\documentclass[journal]{IEEEtran}

\usepackage{epsfig,amsmath,amssymb,epsf,amsthm,scalefnt,multirow,bm,mathtools,stfloats,stmaryrd}
\usepackage{xcolor}
\usepackage{float}
\usepackage{cite}
\usepackage{psfrag}
\usepackage{algorithm}
\usepackage{algpseudocode}

\newcommand{\bl}{\big(}
\newcommand{\br}{\big)}

\newtheorem{lemma}{Lemma}
\newtheorem*{lemma*}{Lemma}

\algrenewcommand\algorithmicrequire{\textbf{Input:}}
\algrenewcommand\algorithmicensure{\textbf{Output:}}
\algdef{SE}[DOWHILE]{Do}{doWhile}{\algorithmicdo}[1]{\algorithmicwhile\ #1}

\begin{document}

\title{Cell-Free MmWave Massive MIMO Systems with Low-Capacity Fronthaul Links and Low-Resolution ADC/DACs}

\author{In-soo Kim, Mehdi Bennis, and Junil Choi
\thanks{Copyright (c) 2015 IEEE. Personal use of this material is permitted. However, permission to use this material for any other purposes must be obtained from the IEEE by sending a request to pubs-permissions@ieee.org.}
\thanks{I. Kim and J. Choi are with the School of Electrical Engineering, KAIST, Daejeon, Korea (e-mail: \{insookim; junil\}@kaist.ac.kr).}
\thanks{M. Bennis is with the Centre for Wireless Communications, University of Oulu, Oulu, Finland (e-mail: mehdi.bennis@oulu.fi).}
\thanks{This work was supported in part by the MSIT (Ministry of Science and ICT), Korea, under the ITRC (Information Technology Research Center) support program (IITP-2020-0-01787) supervised by the IITP (Institute of Information \& Communications Technology Planning \& Evaluation), in part by the Institute of Information \& Communications Technology Planning \& Evaluation (IITP) grant funded by the Korea government (MSIT) (No. 2021-000269, Development of sub-THz band wireless transmission and access core technology for 6G Tbps data rate), and in part by the Basic Science Research Program through the National Research Foundation of Korea (NRF) funded by the Ministry of Education (No. 2021R1A6A3A13045215).}}

\maketitle

\begin{abstract}
In this paper, we consider the uplink channel estimation phase and downlink data transmission phase of cell-free millimeter wave (mmWave) massive multiple-input multiple-output (MIMO) systems with low-capacity fronthaul links and low-resolution analog-to-digital converters/digital-to-analog converters (ADC/DACs). In cell-free massive MIMO, a control unit dictates the baseband processing at a geographical scale, while the base stations communicate with the control unit through fronthaul links. Unlike most of previous works in cell-free massive MIMO with finite-capacity fronthaul links, we consider the general case where the fronthaul capacity and ADC/DAC resolution are not necessarily the same. In particular, the fronthaul compression and ADC/DAC quantization occur independently where each one is modeled based on the information theoretic argument and additive quantization noise model (AQNM). Then, we address the codebook design problem that aims to minimize the channel estimation error for the independent and identically distributed (i.i.d.) and colored compression noise cases. Also, we propose an alternating optimization (AO) method to tackle the max-min fairness problem. In essence, the AO method alternates between two subproblems that correspond to the power allocation and codebook design problems. The AO method proposed for the zero-forcing (ZF) precoder is guaranteed to converge, whereas the one for the maximum ratio transmission (MRT) precoder has no such guarantee. Finally, the performance of the proposed schemes is evaluated by the simulation results in terms of both energy and spectral efficiency. The numerical results show that the proposed scheme for the ZF precoder yields spectral and energy efficiency 28\% and 15\% higher than that of the best baseline.
\end{abstract}

\begin{IEEEkeywords}
Max-min fairness, cell-free massive multiple-input multiple-output (MIMO), fronthaul compression, quantization, analog-to-digital converter (ADC), digital-to-analog converter (DAC).
\end{IEEEkeywords}

\section{Introduction}
\IEEEPARstart{5}{G} wireless communications support the high data rate by densifying the network \cite{6736747}. The inter-cell interference incurred by network densification, however, fueled the demand for a control unit that manages the network at a geographical scale. The idea of centralized wireless communications is known as cell-free massive multiple-input multiple-output (MIMO), which was popularized by \cite{7827017, 7917284}. The gain of cell-free massive MIMO is fully extracted provided that the fronthaul capacity is sufficient for the control unit and base stations to cooperate without much difficulty \cite{8845768}. By communicating at the millimeter wave (mmWave) band in the range of 30-300 GHz, the gain of cell-free massive MIMO is further enhanced to support the unprecedented high data rate \cite{6894453, 7397887}. The power consumption of cell-free massive MIMO, however, may grow without bound as network densification, massive MIMO, and high-capacity fronthaul links demand an untolerable amount of energy \cite{8097026}.

To enhance energy efficiency of cell-free massive MIMO, one possible solution is to accommodate low-capacity fronthaul links that rely on quantization to pass the signal between the control unit and base stations. In practice, quantization is implemented by low-resolution analog-to-digital converters/digital-to-analog converters (ADC/DACs) deployed at the radio frequency (RF) chains of the base stations. Recent works in quantizer-based cell-free massive MIMO include \cite{8756286, 8781848, 9212395, 8756265, 9123382} where each focused on different aspects of cell-free massive MIMO. In \cite{8756286, 8781848, 9212395}, the uplink of cell-free massive MIMO was analyzed in various aspects where quantization was modeled based on the additive quantization noise model (AQNM). In particular, the max-min fairness problem \cite{8756286}, energy efficiency maximization problem \cite{8781848}, and various quantization strategies \cite{9212395} were addressed to enhance the performance of quantizer-based cell-free massive MIMO. Unlike \cite{8756286, 8781848, 9212395} that focused on the uplink, \cite{8756265, 9123382} considered the downlink of quantizer-based cell-free massive MIMO. Specifically, \cite{8756265} tackled the max-min fairness and ADC/DAC bit allocation problems to maximize spectral efficiency, while \cite{9123382} focused only on the max-min fairness problem. The main shortcoming of \cite{8756286, 8781848, 9212395, 8756265, 9123382} is that the ADC/DAC resolution of the base stations must match the fronthaul capacity, which restricts the system design flexibility.

Another way to implement low-capacity fronthaul links in cell-free massive MIMO is to rely on codebook-based fronthaul compression, which was recently studied in \cite{8891922, 8678745}. In \cite{8891922}, the uplink of cell-free massive MIMO was investigated where the fronthaul compression was modeled based on the rate distortion theory. The main focus was on various compression strategies for the base stations to forward the signal to the control unit subject to the fronthaul capacity constraint, as well as finding the most energy-efficient operating point as a function of the fronthaul capacity. In contrast, \cite{8678745} considered the max-min fairness problem in the downlink of cell-free massive MIMO based on the same fronthaul compression model. The issue of the heuristic solution of \cite{8678745} is that there is no convergence guarantee as the problem under consideration is nonconvex.

To address the issues discussed until now, we consider cell-free mmWave massive MIMO systems with low-capacity fronthaul links and low-resolution ADC/DACs where the fronthaul capacity and ADC/DAC resolution are not necessarily constrained to match. In particular, codebook-based fronthaul compression is performed at fronthaul links, while quantization is carried out by low-resolution ADC/DACs at the base stations. The main contributions of this paper are as follows:
\begin{itemize}
\item We consider the general case where the fronthaul compression and ADC/DAC quantization are performed independently so that the fronthaul capacity and ADC/DAC resolution are allowed to have different values. The fronthaul compression is performed based on the codebook approach, whose compression noise is modeled using the rate-distortion-theory-based information theoretic argument introduced in \cite{6924850, 8891922}. Meanwhile, we model the ADC/DAC quantization noise using the AQNM. As a result, unlike recent works in quantizer-based cell-free massive MIMO \cite{8756286, 8781848, 9212395, 8756265, 9123382}, we examine how energy efficiency of cell-free massive MIMO is affected by the fronthaul capacity and ADC/DAC resolution as investigated in Section \ref{section_4}.
\item The codebook design problem associated with the fronthaul compression is studied in the uplink channel estimation phase where the goal is to minimize the channel estimation error. In particular, we propose a minimum mean squared error-achieving (MMSE-achieving) codebook for the independent and identically distributed (i.i.d.) compression noise case. In the colored compression noise case, we show that the codebook design problem becomes nonconvex with a linear matrix inequality constraint, which is computationally intractable to solve for large systems like cell-free massive MIMO.
\item We propose an alternating optimization (AO) method that solves the max-min fairness problem in the downlink data transmission phase for the maximum ratio transmission (MRT) precoder. Since the max-min fairness problem is nonconvex due to the fronthaul capacity constraint, the heuristic approach developed in \cite{8678745} for the zero-forcing (ZF) precoder with infinite-resolution ADC/DACs is adopted to our case for the MRT precoder. In essense, the scheme proposed in \cite{8678745} consists of a quasilinear optimization problem, while each iteration of our scheme is formulated as a quasiconcave optimization problem. There is no guarantee for the AO method for the MRT precoder to converge, but the issue is not as problematic in practice as the simulation results show.
\item To overcome the limitation of the MRT precoder, we develop a novel AO method to tackle the max-min fairness problem for the ZF precoder. The AO method alternates between the power allocation and codebook design problems, which are nonconvex in nature. Nevertheless, our scheme attains the global optima of the subproblems at each iteration, thereby reaching at the local optimum of the original problem. The proposed scheme outperforms the heuristic approach developed in \cite{8678745} for the ZF precoder as the simulation results reveal.
\end{itemize}

This paper is organized as follows. In Section \ref{section_2}, the uplink channel estimation phase and downlink data transmission phase are discussed. In Section \ref{section_3}, an AO method that solves the max-min fairness problem is proposed for the MRT and ZF precoders. The performance of the codebook design scheme for channel estimation and AO method for the max-min fairness problem is evaluated by the simulation results in Section \ref{section_4}, which is followed by the conclusion in Section \ref{section_5}.

\textbf{Notation:} $a$, $\mathbf{a}$, and $\mathbf{A}$ denote a scalar, vector, and matrix. $\mathbf{0}_{n}$, $\mathbf{0}_{m\times n}$, $\mathbf{I}_{n}$ are an $n\times 1$ zero vector, $m\times n$ zero matrix, and $n\times n$ identity matrix. The vectorization of $\mathbf{A}$ is $\mathrm{vec}\bl\mathbf{A}\br$. The diagonal matrix given by the diagonal elements of $\mathbf{A}$ is $\mathrm{diag}\bl\mathbf{A}\br$. The block-diagonal matrix that contains $\mathbf{A}_{1}, \dots, \mathbf{A}_{n}$ as the block-diagonal elements is $\mathrm{blockdiag}\bl\mathbf{A}_{1}, \dots, \mathbf{A}_{n}\br$. The Kronecker product of $\mathbf{A}$ and $\mathbf{B}$ is $\mathbf{A}\otimes\mathbf{B}$. $\mathbf{A}\succeq\mathbf{B}$ implies that $\mathbf{A}-\mathbf{B}$ is nonnegative definite. $\mathbf{C}_{\mathbf{x}\mathbf{y}}$ denotes $\mathbb{E}\{(\mathbf{x}-\mathbb{E}\{\mathbf{x}\})(\mathbf{y}-\mathbb{E}\{\mathbf{y}\})^{\mathrm{H}}\}$. $I(\mathbf{x}; \mathbf{y})$ is the mutual information of random vectors $\mathbf{x}$ and $\mathbf{y}$. $\delta[n]$ is the Kronecker delta. $\llbracket n\rrbracket$ denotes $\llbracket n\rrbracket=\{1, \dots, n\}$.

\begin{figure}[t]
\centering
\includegraphics[width=1\columnwidth]{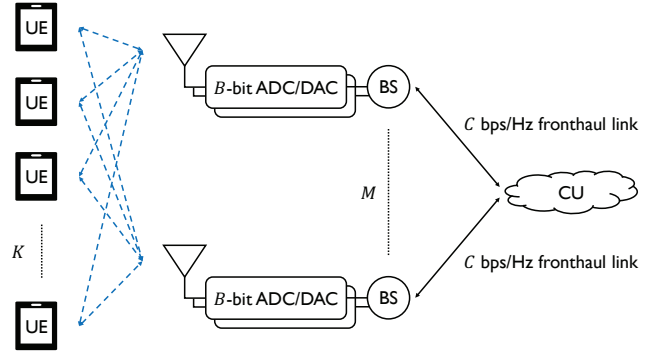}
\caption[caption]{A cell-free mmWave massive MIMO system with $M$ base stations and $K$ single-antenna users where the base stations have a pair of low-resolution ADC/DACs at each RF chain. The baseband processing is carried out at a control unit, which is connected to the base stations that act as radio units with low-capacity fronthaul links.}\label{figure_1}
\end{figure}

\section{System Model}\label{section_2}
Consider a cell-free mmWave massive MIMO system with $M$ base stations and $K$ single-antenna users where each base station has $N$ antennas and $R$ RF chains. The system operates under time-division duplexing (TDD) that exploits the channel reciprocity in the uplink and downlink. The base stations have a pair of $B_{\mathrm{u}}$-bit ADCs ($B_{\mathrm{d}}$-bit DACs) at each RF chain in the uplink (downlink). A control unit performs the baseband processing, while the base stations act as radio units as shown in Fig. \ref{figure_1}. We focus on the low-mobility case where most of the users are pedestrians that move slower than 10 km/h as in \cite{7917284, 9309348, 8756265}.

One coherence block consists of the uplink channel estimation phase and downlink data transmission phase where the uplink proceeds as follows: The received signal quantized by $B_{\mathrm{u}}$-bit ADCs is compressed at the base station, conveyed through a fronthaul link of $C_{\mathrm{u}}$ bps/Hz, and decompressed at the control unit for channel estimation. Likewise, the downlink data transmission phase proceeds by precoding the data signal based on the estimated channel at the control unit, compressing and conveying through a fronthaul link of $C_{\mathrm{d}}$ bps/Hz, and decompressing at the base station for data transmission with $B_{\mathrm{d}}$-bit DACs. In this section, the system model that describes the uplink channel estimation phase and downlink data transmission phase of a cell-free mmWave massive MIMO system with low-capacity fronthaul links and low-resolution ADC/DACs is developed.

\textbf{Remark 1:} Throughout the system model, the base stations and users are assumed to be synchronized almost perfectly, which can be realized by the existing 5G protocols as follows. First, the base stations are synchronized with the neighboring base stations \cite{6760595}. Then, the primary and secondary synchronization signals are broadcasted to the users, which are utilized to pair the clocks of a (master) base station and user \cite{6247449}. After the clocks of the base stations and users are synchronized, however, the signals still can be asynchronously received due to the largely different distances between the base stations and users. In orthogonal frequency-division multiplexing (OFDM) systems, such a misalignment is dealt by controlling the length of the cyclic prefix so as to compensate for the different propagation delays, which results in a quasi-synchronous network. In the 5G standard, the cyclic prefix is long enough to quasi-synchronize the base stations and users within 1 km radius \cite{4287203}.

\subsection{Uplink Channel Estimation Phase}\label{section_2a}
In this subsection, the uplink channel estimation phase is formulated. Then, we propose a codebook design scheme that minimizes the channel estimation error for the i.i.d. compression noise case. In addition, the colored compression noise case is analyzed, which is revealed to be computationally infeasible to solve for large systems like cell-free massive MIMO.

The quantized received signal $\mathbf{Y}_{\mathrm{u}, m}\in\mathbb{C}^{R\times T}$ of the $m$-th base station over the channel estimation phase of length $T\geq K$ is
\begin{align}\label{uplink_quantized_received_signal}
&\mathbf{Y}_{\mathrm{u}, m}=\mathrm{Q}_{\mathrm{u}}\left(\mathbf{W}_{m}^{\mathrm{H}}\left(\sum_{k=1}^{K}\mathbf{h}_{m, k}\mathbf{x}_{\mathrm{u}, k}^{\mathrm{H}}+\mathbf{N}_{\mathrm{u}, m}\right)\right)=\notag\\
&\mathrm{Q}_{\mathrm{u}}\left(\mathbf{W}_{m}^{\mathrm{H}}\underbrace{\begin{bmatrix}\mathbf{h}_{m, 1}&\cdots&\mathbf{h}_{m, K}\end{bmatrix}}_{=\mathbf{H}_{m}}\underbrace{\begin{bmatrix}\mathbf{x}_{\mathrm{u}, 1}^{\mathrm{H}}\\\vdots\\\mathbf{x}_{\mathrm{u}, K}^{\mathrm{H}}\end{bmatrix}}_{=\mathbf{X}_{\mathrm{u}}^{\mathrm{H}}}+\underbrace{\mathbf{W}_{m}^{\mathrm{H}}\mathbf{N}_{\mathrm{u}, m}}_{=\mathbf{V}_{\mathrm{u}, m}}\right)
\end{align}
where $\mathbf{h}_{m, k}\sim\mathcal{CN}\bl\mathbf{0}_{N}, \mathbf{C}_{\mathbf{h}_{m, k}}\br$ is the channel between the $m$-th base station and $k$-th user, $\mathbf{x}_{\mathrm{u}, k}\in\mathbb{C}^{T}$ is the training signal of the $k$-th user that satisfies $\mathbf{x}_{\mathrm{u}, k}^{\mathrm{H}}\mathbf{x}_{\mathrm{u}, \ell}=TP_{\mathrm{u}}\delta[k-\ell]$ so as to prevent pilot contamination,\footnote{As mentioned earlier, the users are assumed to move slowly, which is a commonly adopted scenario in cell-free massive MIMO. In the low-mobility case, the coherence block length is long enough to assign orthogonal training signals to the users \cite{7917284, 9309348, 8756265}.} $\mathbf{N}_{\mathrm{u}, m}\in\mathbb{C}^{N\times T}$ is the additive white Gaussian noise (AWGN) of the $m$-th base station with i.i.d. $\mathcal{CN}(0, \sigma_{\mathrm{u}}^{2})$ elements, and $\mathbf{W}_{m}\in\mathbb{C}^{N\times R}$ is the RF combiner of the $m$-th base station with unit-modulus elements. In practice, the phase shifters of the analog beamformers cannot be switched instantaneously due to the hardware constraint imposed on the phase shifters \cite{7961152, 8306126, 8323164}. As a result, analog beamformers that depend on the slowly-varying second-order channel statistics are commonly used in the mmWave hybrid massive MIMO literatures \cite{7919262, 7833195, 7908940, 8678745}, which is referred to as statistical beamforming. Therefore, we configure the RF combiner and RF precoder using the second-order channel statistics that remains constant throughout the uplink channel estimation phase and downlink data transmission phase, which implies that the RF combiner and RF precoder are identical. The quantization function $\mathrm{Q}_{\mathrm{u}}(\cdot)$ represents $B_{\mathrm{u}}$-bit ADCs at the RF chains of the base stations that quantizes the real and imaginary parts of each element with $2^{B_{\mathrm{u}}}$ quantization points that achieve the MMSE on Gaussian signals \cite{1057548}. The channel consists of $L_{m, k}$ paths that add up to
\begin{equation}\label{channel}
\mathbf{h}_{m, k}=\sum_{\ell=1}^{L_{m, k}}\alpha_{m, k, \ell}\mathbf{a}(\theta_{m, k, \ell}, \phi_{m, k, \ell})
\end{equation}
where $\alpha_{m, k, \ell}\sim\mathcal{CN}(0, \sigma_{m, k, \ell}^{2})$ is the $\ell$-th path gain, $\theta_{m, k, \ell}\in[-\pi, \pi]$ is the $\ell$-th azimuth angle-of-arrival (AoA), $\phi_{m, k, \ell}\in[0, \pi]$ is the $\ell$-th zenith angle-of-arrival (ZoA), and $\mathbf{a}(\theta, \phi)\in\mathbb{C}^{N}$ is the array response vector of the base station. The path gains vary at each coherence block, while the AoAs and ZoAs remain constant over multiple coherence blocks. Therefore, the long-term channel state information (CSI) $\mathbf{C}_{\mathbf{h}_{m, k}}$ defined as
\begin{equation}
\mathbf{C}_{\mathbf{h}_{m, k}}=\sum_{\ell=1}^{L_{m, k}}\sigma_{m, k, \ell}^{2}\mathbf{a}(\theta_{m, k, \ell}, \phi_{m, k, \ell})\mathbf{a}(\theta_{m, k, \ell}, \phi_{m, k, \ell})^{\mathrm{H}}
\end{equation}
is assumed to be known a priori to the control unit. Before moving on, we emphasize that the RF combiner and RF precoder are configured based on the long-term CSI in the simulation results.

To facilitate the analysis, let us vectorize \eqref{uplink_quantized_received_signal} as
\begin{equation}\label{uplink_vectorized_quantized_received_signal}
\underbrace{\mathrm{vec}\bl\mathbf{Y}_{\mathrm{u}, m}\br}_{=\mathbf{y}_{\mathrm{u}, m}}=\mathrm{Q}_{\mathrm{u}}\bl\bl\underbrace{\mathbf{X}_{\mathrm{u}}^{*}\otimes\mathbf{W}_{m}^{\mathrm{H}}}_{=\mathbf{\Psi}_{m}}\br\underbrace{\mathrm{vec}\bl\mathbf{H}_{m}\br}_{=\mathbf{h}_{m}}+\underbrace{\mathrm{vec}\bl\mathbf{V}_{\mathrm{u}, m}\br}_{=\mathbf{v}_{\mathrm{u}, m}}\br
\end{equation}
where $\mathbf{h}_{m}\sim\mathcal{CN}\bl\mathbf{0}_{NK}, \mathbf{C}_{\mathbf{h}_{m}}\br$ with covariance matrix $\mathbf{C}_{\mathbf{h}_{m}}=\mathrm{blockdiag}\bl\mathbf{C}_{\mathbf{h}_{m, 1}}, \dots, \mathbf{C}_{\mathbf{h}_{m, K}}\br$ and $\mathbf{v}_{\mathrm{u}, m}\sim\mathcal{CN}\bl\mathbf{0}_{RT}, \sigma_{\mathrm{u}}^{2}\mathbf{I}_{T}\otimes\mathbf{W}_{m}^{\mathrm{H}}\mathbf{W}_{m}\br$. Since $\mathbf{h}_{m}$ and $\mathbf{v}_{\mathrm{u}, m}$ are Gaussian, the AQNM \cite{9174405} is applicable to \eqref{uplink_vectorized_quantized_received_signal}, which is a variant of the Bussgang theorem \cite{bussgang1952crosscorrelation} that linearizes the nonlinear distortion on Gaussian signals as
\begin{align}
\mathbf{y}_{\mathrm{u}, m}&=(1-\rho_{\mathrm{u}})\bl\mathbf{\Psi}_{m}\mathbf{h}_{m}+\mathbf{v}_{\mathrm{u}, m}\br+\mathbf{q}_{\mathrm{u}, m}\notag\\
                          &=(1-\rho_{\mathrm{u}})\mathbf{\Psi}_{m}\mathbf{h}_{m}+(1-\rho_{\mathrm{u}})\mathbf{v}_{\mathrm{u}, m}+\mathbf{q}_{\mathrm{u}, m}
\end{align}
where $\rho_{\mathrm{u}}<1$ is the quantization distortion factor associated with the ADC resolution and $\mathbf{q}_{\mathrm{u}, m}\in\mathbb{C}^{RT}$ is the quantization noise uncorrelated with the random variables in the ADC input $\mathbf{h}_{m}$ and $\mathbf{v}_{\mathrm{u}, m}$. The exact values of $\rho_{\mathrm{u}}$ are provided in \cite{1057548}, which can be approximated as $\rho_{\mathrm{u}}\approx\pi\sqrt{3}/2\cdot 2^{-2B_{\mathrm{u}}}$ when $B_{\mathrm{u}}\gg 1$ \cite{9174405}. The covariance matrix of $\mathbf{q}_{\mathrm{u}, m}$ is \cite{9174405}
\begin{equation}
\mathbf{C}_{\mathbf{q}_{\mathrm{u}, m}}=\rho_{\mathrm{u}}(1-\rho_{\mathrm{u}})\mathrm{diag}\bl\mathbf{\Psi}_{m}\mathbf{C}_{\mathbf{h}_{m}}\mathbf{\Psi}_{m}^{\mathrm{H}}+\sigma_{\mathrm{u}}^{2}\mathbf{I}_{T}\otimes\mathbf{W}_{m}^{\mathrm{H}}\mathbf{W}_{m}\br,
\end{equation}
but the distribution of $\mathbf{q}_{\mathrm{u}, m}$ is unknown and not necessarily Gaussian.

After the reception of $\mathbf{y}_{\mathrm{u}, m}$, the fronthaul link of $C_{\mathrm{u}}$ bps/Hz conveys $\mathbf{y}_{\mathrm{u}, m}$ to the control unit by mapping $\mathbf{y}_{\mathrm{u}, m}$ to some codeword $\hat{\mathbf{y}}_{\mathrm{u}, m}$ at the base station, conveying through the fronthaul link in the form of bits, and demapping the bits to $\hat{\mathbf{y}}_{\mathrm{u}, m}$ at the control unit for channel estimation. The codewords reach the control unit without error as long as the codebook size is smaller than $2^{TC_{\mathrm{u}}}$ because the fronthaul link can deliver $TC_{\mathrm{u}}$ bits per coherence block.\footnote{The unit of $T$ is channel use.}

To model the distortion that comes from the compression-decompression between $\mathbf{y}_{\mathrm{u}, m}$ and $\hat{\mathbf{y}}_{\mathrm{u}, m}$, the information theoretic argument presented in \cite{6924850, 8891922} is adopted.\footnote{The derivation of the fronthaul compression model is based on the rate-distortion theory, and the interested reader is referred to \cite{8891922}.} In particular, \cite{6924850} guarantees the existence of a codebook that makes the joint distribution of $\mathbf{y}_{\mathrm{u}, m}$ and $\hat{\mathbf{y}}_{\mathrm{u}, m}$ close to
\begin{align}\label{yum}
\hat{\mathbf{y}}_{\mathrm{u}, m}&=\mathbf{y}_{\mathrm{u}, m}+\mathbf{e}_{\mathrm{u}, m}\notag\\
                                &=(1-\rho_{\mathrm{u}})\mathbf{\Psi}_{m}\mathbf{h}_{m}+(1-\rho_{\mathrm{u}})\mathbf{v}_{\mathrm{u}, m}+\mathbf{q}_{\mathrm{u}, m}+\mathbf{e}_{\mathrm{u}, m}
\end{align}
where $\mathbf{e}_{\mathrm{u}, m}\sim\mathcal{CN}\bl\mathbf{0}_{RT}, \sigma_{\mathrm{u}, m}^{2}\mathbf{I}_{RT}\br$ is the compression noise independent of $\mathbf{y}_{\mathrm{u}, m}$ with $\sigma_{\mathrm{u}, m}\geq 0$ determining the shape of the codebook. The codebook must satisfy $I(\mathbf{y}_{\mathrm{u}, m}; \hat{\mathbf{y}}_{\mathrm{u}, m})\leq TC_{\mathrm{u}}$ for the codewords to pass the fronthaul link without error, but the closed-form expression of $I(\mathbf{y}_{\mathrm{u}, m}; \hat{\mathbf{y}}_{\mathrm{u}, m})$ cannot be obtained due to the non-Gaussianity of $\mathbf{q}_{\mathrm{u}, m}$. Therefore, the upper bound of $I(\mathbf{y}_{\mathrm{u}, m}; \hat{\mathbf{y}}_{\mathrm{u}, m})$ attained by treating $\mathbf{q}_{\mathrm{u}, m}$ as Gaussian is considered,\footnote{Among the random vectors that have the same covariance matrix, Gaussian is the entropy maximizer \cite{cover1999elements}.} which constrains $\sigma_{\mathrm{u}, m}$ to satisfy
\begin{align}\label{cum}
&I(\mathbf{y}_{\mathrm{u}, m}; \hat{\mathbf{y}}_{\mathrm{u}, m})\leq\notag\\
&C_{\mathrm{u}, m}=\log_{2}\mathrm{det}\left(\mathbf{I}_{RT}+\frac{1}{\sigma_{\mathrm{u}, m}^{2}}\mathbf{C}_{\mathbf{y}_{\mathrm{u}, m}}\right)\leq TC_{\mathrm{u}}
\end{align}
where
\begin{align}
\mathbf{C}_{\mathbf{y}_{\mathrm{u}, m}}=&(1-\rho_{\mathrm{u}})^{2}\mathbf{\Psi}_{m}\mathbf{C}_{\mathbf{h}_{m}}\mathbf{\Psi}_{m}^{\mathrm{H}}+\notag\\
                                        &(1-\rho_{\mathrm{u}})^{2}\sigma_{\mathrm{u}}^{2}\mathbf{I}_{T}\otimes\mathbf{W}_{m}^{\mathrm{H}}\mathbf{W}_{m}+\mathbf{C}_{\mathbf{q}_{\mathrm{u}, m}}.
\end{align}
The choice of $\sigma_{\mathrm{u}, m}$ depends on the criterion under consideration, and minimizing the channel estimation error is of our main interest. The optimization over $\sigma_{\mathrm{u}, m}$ has the interpretation of designing a codebook that minimizes the channel estimation error parameterized by $\sigma_{\mathrm{u}, m}$. The codebook size cannot be arbitrarily large and is constrained to the fronthaul capacity as in \eqref{cum}.

As $\hat{\mathbf{y}}_{\mathrm{u}, m}$ reaches the control unit, the control unit performs linear MMSE (LMMSE) estimation of $\mathbf{h}_{m}$ as
\begin{equation}
\hat{\mathbf{h}}_{m}=\begin{bmatrix}\hat{\mathbf{h}}_{m, 1}\\\vdots\\\hat{\mathbf{h}}_{m, K}\end{bmatrix}=\mathbf{C}_{\mathbf{h}_{m}\hat{\mathbf{y}}_{\mathrm{u}, m}}\mathbf{C}_{\hat{\mathbf{y}}_{\mathrm{u}, m}}^{-1}\hat{\mathbf{y}}_{\mathrm{u}, m}
\end{equation}
with the covariance matrices of the channel estimate $\hat{\mathbf{h}}_{m}$ and channel estimation error $\tilde{\mathbf{h}}_{m}=\mathbf{h}_{m}-\hat{\mathbf{h}}_{m}$ being
\begin{align}
  \mathbf{C}_{\hat{\mathbf{h}}_{m}}&=\mathrm{blockdiag}\bl\mathbf{C}_{\hat{\mathbf{h}}_{m, 1}}, \dots, \mathbf{C}_{\hat{\mathbf{h}}_{m, K}}\br\notag\\
                                   &=\mathbf{C}_{\mathbf{h}_{m}\hat{\mathbf{y}}_{\mathrm{u}, m}}\mathbf{C}_{\hat{\mathbf{y}}_{\mathrm{u}, m}}^{-1}\mathbf{C}_{\mathbf{h}_{m}\hat{\mathbf{y}}_{\mathrm{u}, m}}^{\mathrm{H}},\\
\mathbf{C}_{\tilde{\mathbf{h}}_{m}}&=\mathrm{blockdiag}\bl\mathbf{C}_{\tilde{\mathbf{h}}_{m, 1}}, \dots, \mathbf{C}_{\tilde{\mathbf{h}}_{m, K}}\br\notag\\
                                   &=\mathbf{C}_{\mathbf{h}_{m}}-\mathbf{C}_{\hat{\mathbf{h}}_{m}}
\end{align}
where
\begin{align}
&\mathbf{C}_{\mathbf{h}_{m}\hat{\mathbf{y}}_{\mathrm{u}, m}}=(1-\rho_{\mathrm{u}})\mathbf{C}_{\mathbf{h}_{m}}\mathbf{\Psi}_{m}^{\mathrm{H}},\\
&\mathbf{C}_{\hat{\mathbf{y}}_{\mathrm{u}, m}}=\mathbf{C}_{\mathbf{y}_{\mathrm{u}, m}}+\sigma_{\mathrm{u}, m}^{2}\mathbf{I}_{RT}.
\end{align}
The block-diagonal structure of $\mathbf{C}_{\hat{\mathbf{h}}_{m}}$ and $\mathbf{C}_{\tilde{\mathbf{h}}_{m}}$ results from the orthogonality of the training signals that prevents pilot contamination. The distributions of $\hat{\mathbf{h}}_{m}$ and $\tilde{\mathbf{h}}_{m}$, however, are unknown because $\mathbf{q}_{\mathrm{u}, m}$ is non-Gaussian.

To choose $\sigma_{\mathrm{u}, m}$ that minimizes the channel estimation error among the ones that satisfy \eqref{cum}, let us solve
\begin{alignat}{2}\label{p1}
&\underset{\sigma_{\mathrm{u}, m}\geq 0}{\mathrm{min}}\enspace&&\mathrm{tr}\bl\mathbf{C}_{\tilde{\mathbf{h}}_{m}}\br\notag\\
&\mathrm{subject}\ \mathrm{to}\enspace                        &&C_{\mathrm{u}, m}\leq TC_{\mathrm{u}}.\tag{P1}
\end{alignat}
To solve \eqref{p1}, note that minimizing $\mathrm{tr}\bl\mathbf{C}_{\tilde{\mathbf{h}}_{m}}\br$ is equivalent to maximizing
\begin{equation}\label{trchmhat}
\mathrm{tr}\bl\mathbf{C}_{\hat{\mathbf{h}}_{m}}\br=\mathrm{tr}\bl\mathbf{C}_{\mathbf{h}_{m}\hat{\mathbf{y}}_{\mathrm{u}, m}}\bl\mathbf{C}_{\mathbf{y}_{\mathrm{u}, m}}+\sigma_{\mathrm{u}, m}^{2}\mathbf{I}_{RT}\br^{-1}\mathbf{C}_{\mathbf{h}_{m}\hat{\mathbf{y}}_{\mathrm{u}, m}}^{\mathrm{H}}\br.
\end{equation}
Since $\mathrm{tr}\bl\mathbf{C}_{\hat{\mathbf{h}}_{m}}\br$ and $C_{\mathrm{u}, m}$ are nonincreasing functions of $\sigma_{\mathrm{u}, m}$ as evident from \eqref{cum} and \eqref{trchmhat}, the solution of \eqref{p1} is the one that satisfies the inequality constraint with equality. The solution $\sigma_{\mathrm{u}, m}$ of $C_{\mathrm{u}, m}=TC_{\mathrm{u}}$ can be found numerically using root-finding algorithms that leverage the differentiability of $C_{\mathrm{u}, m}$ with respect to $\sigma_{\mathrm{u}, m}$ like Newton's method.

\textbf{Remark 2:} The compression noise with i.i.d. elements where $\mathbf{e}_{\mathrm{u}, m}\sim\mathcal{CN}\bl\mathbf{0}_{RT}, \sigma_{\mathrm{u}, m}^{2}\mathbf{I}_{RT}\br$ was of our main interest. In general, however, the compression noise can be colored with correlated elements, which makes the covariance matrix $\mathbf{C}_{\mathbf{e}_{\mathrm{u}, m}}\succeq\mathbf{0}_{2RT\times 2RT}$ not diagonal anymore. Since allowing $\mathbf{C}_{\mathbf{e}_{\mathrm{u}, m}}$ to have nonzero off-diagonal elements provides more degrees of freedom, we can design a better codebook. The fact that finding $\mathbf{C}_{\mathbf{e}_{\mathrm{u}, m}}$ that minimizes the channel estimation error is a nonconvex optimization problem, however, makes such a codebook infeasible in practice. In particular, $\mathrm{tr}\bl\mathbf{C}_{\hat{\mathbf{h}}_{m}}\br$ and $C_{\mathrm{u}, m}$ becomes
\begin{align*}
&\mathrm{tr}\bl\mathbf{C}_{\hat{\mathbf{h}}_{m}}\br=\mathrm{tr}\bl\mathbf{C}_{\mathbf{h}_{m}\hat{\mathbf{y}}_{\mathrm{u}, m}}\bl\mathbf{C}_{\mathbf{y}_{\mathrm{u}, m}}+\mathbf{C}_{\mathbf{e}_{\mathrm{u}, m}}\br^{-1}\mathbf{C}_{\mathbf{h}_{m}\hat{\mathbf{y}}_{\mathrm{u}, m}}^{\mathrm{H}}\br,\\
&C_{\mathrm{u}, m}=\log_{2}\mathrm{det}\bl\mathbf{I}_{RT}+\mathbf{C}_{\mathbf{e}_{\mathrm{u}, m}}^{-1}\mathbf{C}_{\mathbf{y}_{\mathrm{u}, m}}\br,
\end{align*}
and the optimization problem is given by
\begin{alignat}{2}\label{p2}
&\underset{\mathbf{C}_{\mathbf{e}_{\mathrm{u}, m}}\succeq\mathbf{0}_{RT\times RT}}{\mathrm{max}}\enspace&&\mathrm{tr}\bl\mathbf{C}_{\hat{\mathbf{h}}_{m}}\br\notag\\
&\mathrm{subject}\ \mathrm{to}\enspace                                                                  &&C_{\mathrm{u}, m}\leq TC_{\mathrm{u}}\tag{P2}
\end{alignat}
analogous to the i.i.d. case. To see why \eqref{p2} is nonconvex, let us analyze the objective function and constraint one by one.

The constraint can be recast as
\begin{align*}
&\log_{2}\mathrm{det}\bl\mathbf{I}_{RT}+\mathbf{C}_{\mathbf{e}_{\mathrm{u}, m}}^{-1}\mathbf{C}_{\mathbf{y}_{\mathrm{u}, m}}\br=\\
&\log_{2}\mathrm{det}\bl\mathbf{I}_{RT}+\mathbf{C}_{\mathbf{y}_{\mathrm{u}, m}}^{1/2}\mathbf{C}_{\mathbf{e}_{\mathrm{u}, m}}^{-1}\mathbf{C}_{\mathbf{y}_{\mathrm{u}, m}}^{1/2}\br\leq TC_{\mathrm{u}},
\intertext{\centering$\overset{(a)}{\iff}$}
&\log_{2}\mathrm{det}\bl\mathbf{Z}^{-1}\br\leq TC_{\mathrm{u}},\\
&\mathbf{Z}\preceq\bl\mathbf{I}_{RT}+\mathbf{C}_{\mathbf{y}_{\mathrm{u}, m}}^{1/2}\mathbf{C}_{\mathbf{e}_{\mathrm{u}, m}}^{-1}\mathbf{C}_{\mathbf{y}_{\mathrm{u}, m}}^{1/2}\br^{-1}=\\
&\mathbf{I}_{RT}-\mathbf{C}_{\mathbf{y}_{\mathrm{u}, m}}^{1/2}\bl\mathbf{C}_{\mathbf{e}_{\mathrm{u}, m}}+\mathbf{C}_{\mathbf{y}_{\mathrm{u}, m}}\br^{-1}\mathbf{C}_{\mathbf{y}_{\mathrm{u}, m}}^{1/2},
\intertext{\centering$\overset{(b)}{\iff}$}
&\log_{2}\mathrm{det}\bl\mathbf{Z}^{-1}\br\leq TC_{\mathrm{u}},\\
&\begin{bmatrix}\mathbf{I}_{RT}-\mathbf{Z}&\mathbf{C}_{\mathbf{y}_{\mathrm{u}, m}}^{1/2}\\\mathbf{C}_{\mathbf{y}_{\mathrm{u}, m}}^{1/2}&\mathbf{C}_{\mathbf{e}_{\mathrm{u}, m}}+\mathbf{C}_{\mathbf{y}_{\mathrm{u}, m}}\end{bmatrix}\succeq\mathbf{0}_{2RT\times 2RT}
\end{align*}
where (a) comes from introducing $\mathbf{Z}$ that upper bounds the constraint function as $C_{\mathrm{u}, m}\leq\log_{2}\mathrm{det}\bl\mathbf{Z}^{-1}\br$ in conjunction with the matrix inversion lemma \cite{woodbury1950inverting}, while (b) is due to the Schur complement. Since the sublevel set of a log-determinant-inverse function and linear matrix inequality define convex sets \cite{boyd2004convex}, we conclude that the constraints in (b) are convex.

Unfortunately, the trace-inverse function in the objective function of \eqref{p2} is convex \cite{boyd2004convex}. Therefore, \eqref{p2} is a nonconvex optimization problem that maximizes a convex objective function over a convex set. There are some algorithms that approximately solve nonconvex optimization problems with objective and constraint functions being sums of convex and concave functions like the disciplined convex-concave programming (DCCP) \cite{7798400}. Nevertheless, solving \eqref{p2} is a difficult task for large systems like cell-free massive MIMO because the linear matrix inequality in (b) is usually dealt with semidefinite programming (SDP), which is computationally burdensome to handle. A similar argument holds in precoder optimization during the downlink data transmission phase. Therefore, the colored case, which is a challenging yet interesting topic, is left for future work.

\textbf{Remark 3:} In this paper, the quantized received signal $\mathbf{y}_{\mathrm{u}, m}$ described by $2TRB_{\mathrm{u}}$ bits is mapped to some codeword $\hat{\mathbf{y}}_{\mathrm{u}, m}$ from a codebook of size $2^{TC_{\mathrm{u}}}$. If the fronthaul capacity is larger than the ADC resolution, an alternative strategy is to forward $2TRB_{\mathrm{u}}$ bits through the fronthaul link instead of conveying the compressed version of the quantized received signal, thereby eliminating the compression noise from \eqref{yum}. Since we can interpret \eqref{yum} without the compression noise as the infinite fronthaul capacity case in our system model, however, we focus on the general case where there is compression noise without loss of generality. The same logic holds in the downlink data transmission phase.

\subsection{Downlink Data Transmission Phase}
In this subsection, the downlink data transmission phase is developed. The quantized transmitted signal $\mathbf{x}_{\mathrm{d}, m}\in\mathbb{C}^{N}$ of the $m$-th base station at each time slot is
\begin{align}\label{downlink_quantized_transmitted_signal}
\mathbf{x}_{\mathrm{d}, m}&=\mathbf{W}_{m}\mathrm{Q}_{\mathrm{d}}\left(\underbrace{\sum_{k=1}^{K}\mathbf{f}_{m, k}\sqrt{p_{m, k}}s_{k}}_{=\mathbf{p}_{\mathrm{d}, m}}+\mathbf{e}_{\mathrm{d}, m}\right)\notag\\
                          &=\mathbf{W}_{m}\mathrm{Q}_{\mathrm{d}}\left(\underbrace{\mathbf{p}_{\mathrm{d}, m}+\mathbf{e}_{\mathrm{d}, m}}_{=\hat{\mathbf{p}}_{\mathrm{d}, m}}\right)
\end{align}
where $\mathbf{f}_{m, k}\in\mathbb{C}^{R}$ and $p_{m, k}\geq 0$ are the linear precoder and power coefficient controlled by the control unit, $s_{k}\sim\mathcal{CN}(0, 1)$ is the data signal to the $k$-th user, $\mathbf{p}_{\mathrm{d}, m}\in\mathbb{C}^{R}$ is the precoded signal transmitted at the $m$-th base station, and $\mathbf{W}_{m}$ is the RF precoder of the $m$-th base station identical to the RF combiner of the uplink channel estimation phase. The quantization function $\mathrm{Q}_{\mathrm{d}}(\cdot)$ models $B_{\mathrm{d}}$-bit DACs that quantize the real and imaginary parts elementwise with $2^{B_{\mathrm{d}}}$ quantization points in \cite{1057548}. The transmit power constraint is $P_{\mathrm{d}, m}=\mathbb{E}\{\|\mathbf{x}_{\mathrm{d}, m}\|^{2}\}\leq P_{\mathrm{d}}$.

The compression noise $\mathbf{e}_{\mathrm{d}, m}\sim\mathcal{CN}\bl\mathbf{0}_{R}, \sigma_{\mathrm{d}, m}^{2}\mathbf{I}_{R}\br$ represents the distortion from the compression of $\mathbf{p}_{\mathrm{d}, m}$ in the fronthaul link of $C_{\mathrm{d}}$ bps/Hz. In particular, $\mathbf{p}_{\mathrm{d}, m}$ reaches the base station by mapping $\mathbf{p}_{\mathrm{d}, m}$ to some codeword $\hat{\mathbf{p}}_{\mathrm{d}, m}$ at the control unit, conveying through the fronthaul link in bits, and demapping the bits to $\hat{\mathbf{p}}_{\mathrm{d}, m}$ at the base station. Similar to the uplink, $\mathbf{e}_{\mathrm{d}, m}$ is independent of $\mathbf{p}_{\mathrm{d}, m}$ and $\sigma_{\mathrm{d}, m}\geq 0$ determines the shape of the codebook. The existence of a codebook that makes $\mathbf{e}_{\mathrm{d}, m}$ close to $\mathcal{CN}\bl\mathbf{0}_{R}, \sigma_{\mathrm{d}, m}^{2}\mathbf{I}_{R}\br$ is guaranteed by the information theoretic argument proposed in \cite{6924850, 8891922} as long as the codebook size is smaller than $2^{C_{\mathrm{d}}}$, which defines the feasible set of $\sigma_{\mathrm{d}, m}$ as the ones that satisfy
\begin{align}\label{cdm}
C_{\mathrm{d}, m}&=I(\mathbf{p}_{\mathrm{d}, m}; \hat{\mathbf{p}}_{\mathrm{d}, m})\notag\\
                 &=\log_{2}\mathrm{det}\left(\mathbf{I}_{R}+\frac{1}{\sigma_{\mathrm{d}, m}^{2}}\sum_{k=1}^{K}\mathbf{f}_{m, k}\mathbf{f}_{m, k}^{\mathrm{H}}p_{m, k}\right)\leq C_{\mathrm{d}}.
\end{align}
Before moving on, note that the closed-form expression of $I(\mathbf{p}_{\mathrm{d}, m}; \hat{\mathbf{p}}_{\mathrm{d}, m})$ is obtained in the downlink data transmission phase unlike the uplink that deals with the upper bound of $I(\mathbf{y}_{\mathrm{u}, m}; \hat{\mathbf{y}}_{\mathrm{u}, m})$. The discrepancy between the uplink channel estimation phase and downlink data transmission phase comes from the direction of the signal flow: Is the signal compressed first or quantized first?

Like the uplink, the Gaussianity of $s_{k}$ and $\mathbf{e}_{\mathrm{d}, m}$ allows the AQNM \cite{9174405} to linearize \eqref{downlink_quantized_transmitted_signal} as
\begin{align}
&\mathbf{x}_{\mathrm{d}, m}=\notag\\
&\mathbf{W}_{m}\left((1-\rho_{\mathrm{d}})\left(\sum_{k=1}^{K}\mathbf{f}_{m, k}\sqrt{p_{m, k}}s_{k}+\mathbf{e}_{\mathrm{d}, m}\right)+\mathbf{q}_{\mathrm{d}, m}\right)=\notag\\
&\mathbf{W}_{m}\left(\sum_{k=1}^{K}(1-\rho_{\mathrm{d}})\mathbf{f}_{m, k}\sqrt{p_{m, k}}s_{k}+(1-\rho_{\mathrm{d}})\mathbf{e}_{\mathrm{d}, m}+\mathbf{q}_{\mathrm{d}, m}\right)
\end{align}
where $\rho_{\mathrm{d}}<1$ is the quantization distortion factor associated with the DAC resolution, whose exact values are provided in \cite{1057548} and can be approximated as $\rho_{\mathrm{d}}\approx\pi\sqrt{3}/2\cdot 2^{-2B_{\mathrm{d}}}$ when $B_{\mathrm{d}}\gg 1$ \cite{9174405}, and $\mathbf{q}_{\mathrm{d}, m}\in\mathbb{C}^{R}$ is the quantization noise uncorrelated with the random variables in the DAC input $s_{k}$ and $\mathbf{e}_{\mathrm{d}, m}$. Again, the covariance matrix of $\mathbf{q}_{\mathrm{d}, m}$ is \cite{9174405}
\begin{equation}
\mathbf{C}_{\mathbf{q}_{\mathrm{d}, m}}=\rho_{\mathrm{d}}(1-\rho_{\mathrm{d}})\mathrm{diag}\left(\sum_{k=1}^{K}\mathbf{f}_{m, k}\mathbf{f}_{m, k}^{\mathrm{H}}p_{m, k}+\sigma_{\mathrm{d}, m}^{2}\mathbf{I}_{R}\right)
\end{equation}
with some unknown distribution.

Using the signal model until now, the received signal $y_{\mathrm{d}, k}\in\mathbb{C}$ of the $k$-th user is expressed as
\begin{align}\label{downlink_received_signal}
&y_{\mathrm{d}, k}=\sum_{m=1}^{M}\mathbf{h}_{m, k}^{\mathrm{H}}\mathbf{x}_{\mathrm{d}, m}+n_{\mathrm{d}, k}=\sum_{m=1}^{M}\underbrace{\mathbf{h}_{m, k}^{\mathrm{H}}\mathbf{W}_{m}}_{=\mathbf{g}_{m, k}^{\mathrm{H}}}\notag\\
&\left(\sum_{i=1}^{K}(1-\rho_{\mathrm{d}})\mathbf{f}_{m, i}\sqrt{p_{m, i}}s_{i}+\underbrace{(1-\rho_{\mathrm{d}})\mathbf{e}_{\mathrm{d}, m}+\mathbf{q}_{\mathrm{d}, m}}_{=\mathbf{v}_{\mathrm{d}, m}}\right)+n_{\mathrm{d}, k}
\end{align}
where $n_{\mathrm{d}, k}\sim\mathcal{CN}(0, \sigma_{\mathrm{d}}^{2})$ is the AWGN of the $k$-th user. Also, the channel estimate $\hat{\mathbf{g}}_{m, k}=\mathbf{W}_{m}^{\mathrm{H}}\hat{\mathbf{h}}_{m, k}$ and channel estimation error $\tilde{\mathbf{g}}_{m, k}=\mathbf{g}_{m, k}-\hat{\mathbf{g}}_{m, k}=\mathbf{W}_{m}^{\mathrm{H}}\tilde{\mathbf{h}}_{m, k}$ associated with $\mathbf{g}_{m, k}$ are introduced for notational simplicity in the sequel, whose covariance matrices are
\begin{align}
&\mathbf{C}_{\hat{\mathbf{g}}_{m, k}}=\mathbf{W}_{m}^{\mathrm{H}}\mathbf{C}_{\hat{\mathbf{h}}_{m, k}}\mathbf{W}_{m},\\
&\mathbf{C}_{\tilde{\mathbf{g}}_{m, k}}=\mathbf{W}_{m}^{\mathrm{H}}\mathbf{C}_{\tilde{\mathbf{h}}_{m, k}}\mathbf{W}_{m}
\end{align}
with some unknown distributions. Likewise, $\mathbf{v}_{\mathrm{d}, m}$ has some unknown distribution with covariance matrix
\begin{align}
\mathbf{C}_{\mathbf{v}_{\mathrm{d}, m}}=&\rho_{\mathrm{d}}(1-\rho_{\mathrm{d}})\mathrm{diag}\left(\sum_{k=1}^{K}\mathbf{f}_{m, k}\mathbf{f}_{m, k}^{\mathrm{H}}p_{m, k}\right)+\notag\\
                                        &(1-\rho_{\mathrm{d}})\sigma_{\mathrm{d}, m}^{2}\mathbf{I}_{R},
\end{align}
and $\mathbf{v}_{\mathrm{d}, m}$ is uncorrelated with $s_{k}$, which is a property inherited from $\mathbf{q}_{\mathrm{d}, m}$.

To distinguish the roles of the control unit and base stations in cell-free massive MIMO, we emphasize that $\mathbf{W}_{m}$ is controlled by the base stations, while $\{\mathbf{f}_{m, k}\}_{\forall m, k}$, $\{p_{m, k}\}_{\forall m, k}$, and $\{\sigma_{\mathrm{d}, m}\}_{\forall m}$ are managed by the control unit. In addition, we assume that the control unit adopts conventional linear precoders like the MRT and ZF schemes to set $\{\mathbf{f}_{m, k}\}_{\forall m, k}$, which is a pratical assumption for large systems like cell-free massive MIMO \cite{7827017, 7917284, 8097026}.

Hence, the goal is to optimize $\{p_{m, k}\}_{\forall m, k}$ and $\{\sigma_{\mathrm{d}, m}\}_{\forall m}$ at the control unit to maximize the achievable rates of the users in some sense under the transmit power constraint and \eqref{cdm}. The optimization over $\{p_{m, k}\}_{\forall m, k}$ corresponds to power allocation, while the optimization over $\{\sigma_{\mathrm{d}, m}\}_{\forall m}$ has the operational interpretation of codebook design. In a loose sense, we refer to the optimization over $\{p_{m, k}\}_{\forall m, k}$ and $\{\sigma_{\mathrm{d}, m}\}_{\forall m}$ as precoder optimization.

\section{Max-Min Fairness Problem}\label{section_3}
In this section, two max-min fairness algorithms for linear precoders are proposed. First, the achievable rate lower bound of the users is derived. Then, we tackle the max-min fairness problem for the MRT and ZF precoders. The algorithm for the MRT precoder is an AO method motivated by the heuristic approach proposed in \cite{8678745}, which was originally developed for the ZF precoder with infinite-resolution ADC/DACs. We show that this approach is applicable to our case by showing that each iteration of the AO method is cast as a quasiconcave optimization problem. Lastly, a novel AO method for the ZF precoder is proposed, whose convergence is guaranteed unlike the one for the MRT precoder. Before we proceed, let us introduce the shorthand notations
\begin{align}
&\mathbf{F}_{m}=\begin{bmatrix}\mathbf{f}_{m, 1}&\cdots&\mathbf{f}_{m, K}\end{bmatrix},\\
&\mathbf{P}_{m}=\mathrm{blockdiag}(p_{m, 1}, \dots, p_{m, K}),\\
&\bm{\sigma}=\mathrm{blockdiag}(\sigma_{\mathrm{d}, 1}, \dots, \sigma_{\mathrm{d}, M}).
\end{align}
In the sequel, we use $\mathbf{P}_{m}$ and $\bm{\sigma}$ with $\{p_{m, k}\}_{\forall k}$ and $\{\sigma_{\mathrm{d}, m}\}_{\forall m}$ interchangeably.

\subsection{Achievable Rate Lower Bound}
To derive the achievable rate lower bound of the $k$-th user conditioned on the estimated channel $\hat{\mathbf{g}}=\{\hat{\mathbf{g}}_{m, i}\}_{\forall m, i}$, $y_{\mathrm{d}, k}$ in \eqref{downlink_received_signal} is rearranged as
\begin{align}
&y_{\mathrm{d}, k}=\notag\\
&\underbrace{\sum_{m=1}^{M}(1-\rho_{\mathrm{d}})\hat{\mathbf{g}}_{m, k}^{\mathrm{H}}\mathbf{f}_{m, k}\sqrt{p_{m, k}}s_{k}}_{=T_{0}}+\notag\\
&\underbrace{\sum_{m=1}^{M}(1-\rho_{\mathrm{d}})\tilde{\mathbf{g}}_{m, k}^{\mathrm{H}}\mathbf{f}_{m, k}\sqrt{p_{m, k}}s_{k}}_{=T_{1}}+\notag\\
&\underbrace{\sum_{m=1}^{M}\sum_{i\neq k}(1-\rho_{\mathrm{d}})\mathbf{g}_{m, k}^{\mathrm{H}}\mathbf{f}_{m, i}\sqrt{p_{m, i}}s_{i}}_{=T_{2}}+\underbrace{\sum_{m=1}^{M}\mathbf{g}_{m, k}^{\mathrm{H}}\mathbf{v}_{\mathrm{d}, m}}_{=T_{3}}+n_{\mathrm{d}, k}
\end{align}
where $T_{0}$ is the signal term and $T_{1}+T_{2}+T_{3}$ is the aggregate noise that has non-Gaussian components $\tilde{\mathbf{g}}_{m, k}$ and $\mathbf{\mathbf{v}}_{\mathrm{d}, m}$. Since the closed-form expression of the achievable rate $I(s_{k}; y_{\mathrm{d}, k}|\hat{\mathbf{g}})$ cannot be obtained as $T_{1}$, $T_{2}$, and $T_{3}$ are non-Gaussian, the achievable rate lower bound that treats the aggregate noise as Gaussian \cite{959289} is of our main interest, namely \cite{8845768}
\begin{equation}\label{lower_bound}
\log_{2}\bl 1+\mathrm{SINR}_{k}\bl\{\mathbf{P}_{m}\}_{\forall m}, \bm{\sigma}\br\br\leq I(s_{k}; y_{\mathrm{d}, k}|\hat{\mathbf{g}})
\end{equation}
where
\begin{align}\label{sinrk}
&\mathrm{SINR}_{k}\bl\{\mathbf{P}_{m}\}_{\forall m}, \bm{\sigma}\br=\notag\\
&\frac{\mathbb{E}\{|T_{0}|^{2}|\hat{\mathbf{g}}\}}{\mathbb{E}\{|T_{1}|^{2}|\hat{\mathbf{g}}\}+\mathbb{E}\{|T_{2}|^{2}|\hat{\mathbf{g}}\}+\mathbb{E}\{|T_{3}|^{2}|\hat{\mathbf{g}}\}+\sigma_{\mathrm{d}}^{2}}.
\end{align}
For the lower bound in \eqref{lower_bound} to hold, the aggregate noise must be uncorrelated with $T_{0}$ conditioned on $\hat{\mathbf{g}}$. Since
\begin{equation*}
\mathbb{E}\{s_{k}T_{1}^{*}|\hat{\mathbf{g}}\}=\mathbb{E}\{s_{k}T_{2}^{*}|\hat{\mathbf{g}}\}=\mathbb{E}\{s_{k}T_{3}^{*}|\hat{\mathbf{g}}\}=0
\end{equation*}
as evident from $\mathbb{E}\{\tilde{\mathbf{g}}_{m, k}|\hat{\mathbf{g}}\}=0$, $\mathbb{E}\{s_{k}s_{i}^{*}|\hat{\mathbf{g}}\}=\delta[k-i]$, and $\mathbb{E}\{s_{k}\mathbf{v}_{\mathrm{d}, m}^{\mathrm{H}}|\hat{\mathbf{g}}\}=\mathbf{0}_{R}^{\mathrm{T}}$, the uncorrelatedness of the aggregate noise holds.

The conditional variances of $T_{0}$, $T_{1}$, $T_{2}$, and $T_{3}$ in \eqref{sinrk} are expressed as
\begin{align}
&\mathbb{E}\{|T_{0}|^{2}|\hat{\mathbf{g}}\}=\left|\sum_{m=1}^{M}(1-\rho_{\mathrm{d}})\hat{\mathbf{g}}_{m, k}^{\mathrm{H}}\mathbf{f}_{m, k}\sqrt{p_{m, k}}\right|^{2},\label{vart0}\\
&\mathbb{E}\{|T_{1}|^{2}|\hat{\mathbf{g}}\}=\sum_{m=1}^{M}(1-\rho_{\mathrm{d}})^{2}\mathbf{f}_{m, k}^{\mathrm{H}}\mathbf{C}_{\tilde{\mathbf{g}}_{m, k}}\mathbf{f}_{m, k}p_{m, k},\label{vart1}\\
&\mathbb{E}\{|T_{2}|^{2}|\hat{\mathbf{g}}\}=(1-\rho_{\mathrm{d}})^{2}\sum_{i\neq k}\notag\\
&\left(\left|\sum_{m=1}^{M}\hat{\mathbf{g}}_{m, k}^{\mathrm{H}}\mathbf{f}_{m, i}\sqrt{p_{m, i}}\right|^{2}+\sum_{m=1}^{M}\mathbf{f}_{m, i}^{\mathrm{H}}\mathbf{C}_{\tilde{\mathbf{g}}_{m, k}}\mathbf{f}_{m, i}p_{m, i}\right),\\
&\mathbb{E}\{|T_{3}|^{2}|\hat{\mathbf{g}}\}=(1-\rho_{\mathrm{d}})\sum_{m=1}^{M}\notag\\
&\mathrm{tr}\bl\bl\hat{\mathbf{g}}_{m, k}\hat{\mathbf{g}}_{m, k}^{\mathrm{H}}+\mathbf{C}_{\tilde{\mathbf{g}}_{m, k}}\br\bl\rho_{\mathrm{d}}\mathrm{diag}\bl\mathbf{F}_{m}\mathbf{P}_{m}\mathbf{F}_{m}^{\mathrm{H}}\br+\sigma_{\mathrm{d}, m}^{2}\mathbf{I}_{R}\br\br,\label{vart3}
\end{align}
whose derivation involves some straightforward but tedious algebra and the fact that $\mathbb{E}\{\mathbf{g}_{m, k}\mathbf{g}_{m, k}^{\mathrm{H}}|\hat{\mathbf{g}}\}=\hat{\mathbf{g}}_{m, k}\hat{\mathbf{g}}_{m, k}^{\mathrm{H}}+\mathbf{C}_{\tilde{\mathbf{g}}_{m, k}}$, $\mathbb{E}\{\tilde{\mathbf{g}}_{m, k}\tilde{\mathbf{g}}_{n, k}^{\mathrm{H}}|\hat{\mathbf{g}}\}=\mathbf{C}_{\tilde{\mathbf{g}}_{m, k}}\delta[m-n]$, and $\mathbb{E}\{\mathbf{v}_{\mathrm{d}, m}\mathbf{v}_{\mathrm{d}, n}^{\mathrm{H}}|\hat{\mathbf{g}}\}=\mathbf{C}_{\mathbf{v}_{\mathrm{d}, m}}\delta[m-n]$. The second condition holds because there is no pilot contamination, while the third condition comes from the fact that the fronthaul compression and DAC quantization occur independently across the base stations. Before moving on, let us rewrite $P_{\mathrm{d}, m}=\mathbb{E}\{\|\mathbf{x}_{\mathrm{d}, m}\|^{2}\}$ and $C_{\mathrm{d}, m}$ in \eqref{cdm} using $\mathbf{F}_{m}$ and $\mathbf{P}_{m}$ as
\begin{align}
&P_{\mathrm{d}, m}\bl\mathbf{P}_{m}, \sigma_{\mathrm{d}, m}\br=\notag\\
&\quad(1-\rho_{\mathrm{d}})^{2}\mathrm{tr}\bl\mathbf{W}_{m}\mathbf{F}_{m}\mathbf{P}_{m}\mathbf{F}_{m}^{\mathrm{H}}\mathbf{W}_{m}^{\mathrm{H}}\br+\notag\\
&\quad\rho_{\mathrm{d}}(1-\rho_{\mathrm{d}})\mathrm{tr}\bl\mathbf{W}_{m}\mathrm{diag}\bl\mathbf{F}_{m}\mathbf{P}_{m}\mathbf{F}_{m}^{\mathrm{H}}\br\mathbf{W}_{m}^{\mathrm{H}}\br+\notag\\
&\quad(1-\rho_{\mathrm{d}})\mathrm{tr}\bl\mathbf{W}_{m}\mathbf{W}_{m}^{\mathrm{H}}\br\sigma_{\mathrm{d}, m}^{2},\label{pdm_matrix_form}\\
&C_{\mathrm{d}, m}\bl\mathbf{P}_{m}, \sigma_{\mathrm{d}, m}\br=\log_{2}\mathrm{det}\left(\mathbf{I}_{R}+\frac{1}{\sigma_{\mathrm{d}, m}^{2}}\mathbf{F}_{m}\mathbf{P}_{m}\mathbf{F}_{m}^{\mathrm{H}}\right)\label{cdm_matrix_form}
\end{align}
to facilitate the analysis in the sequel.

\subsection{Max-Min Fairness Algorithm for the MRT Precoder}
In this subsection, we solve the max-min fairness problem for the MRT precoder. In particular, the MRT precoder maximizes the signal power by setting the linear precoder to align with the channel estimate as $\{\mathbf{f}_{m, k}\}_{\forall m, k}=\{\hat{\mathbf{g}}_{m, k}\}_{\forall m, k}$, and the max-min fairness problem is formulated as \cite{7827017, 7917284}
\begin{alignat}{2}\label{p3}
&\underset{\{\mathbf{P}_{m}, \bm{\sigma}\}}{\mathrm{max}}\enspace&&\underset{k\in\llbracket K\rrbracket}{\mathrm{min}}\ \mathrm{SINR}_{\mathrm{MRT}, k}\bl\{\mathbf{P}_{m}\}_{\forall m}, \bm{\sigma}\br\notag\\
&\mathrm{subject}\ \mathrm{to}\enspace                           &&P_{\mathrm{MRT}, \mathrm{d}, m}\bl\mathbf{P}_{m}, \sigma_{\mathrm{d}, m}\br\leq P_{\mathrm{d}},\enspace m\in\llbracket M\rrbracket,\notag\\
&                                                                &&C_{\mathrm{MRT}, \mathrm{d}, m}\bl\mathbf{P}_{m}, \sigma_{\mathrm{d}, m}\br\leq C_{\mathrm{d}},\enspace m\in\llbracket M\rrbracket,\notag\\
&                                                                &&\mathbf{P}_{m}\succeq\mathbf{0}_{K\times K},\enspace m\in\llbracket M\rrbracket,\notag\\
&                                                                &&\bm{\sigma}\succeq\mathbf{0}_{M\times M}\tag{P3}
\end{alignat}
where $\mathrm{SINR}_{\mathrm{MRT}, k}$, $P_{\mathrm{MRT}, \mathrm{d}, m}$, and $C_{\mathrm{MRT}, \mathrm{d}, m}$ are obtained by replacing $\{\mathbf{f}_{m, k}\}_{\forall m, k}$ with $\{\hat{\mathbf{g}}_{m, k}\}_{\forall m, k}$ in \eqref{sinrk}-\eqref{cdm_matrix_form}. Therefore, the overall structure of \eqref{sinrk}-\eqref{cdm_matrix_form} with respect to $\{\mathbf{P}_{m}\}_{\forall m}$ and $\bm{\sigma}$ remains unchanged in the optimization point of view. Since \eqref{p3} is nonconvex, we propose an AO method that alternates between $\{\mathbf{P}_{m}\}_{\forall m}$ and $\bm{\sigma}$. As pointed out earlier, our scheme is inspired by the heuristic approach proposed in \cite{8678745} for the ZF precoder with infinite-resolution ADC/DACs. By proving that \eqref{p3} without the second constraint reduces to a quasiconcave optimization problem, we show that the AO method of \cite{8678745} is applicable to our case.

To proceed, we address the optimization over $\{\mathbf{P}_{m}\}_{\forall m}$ first. Since $C_{\mathrm{MRT}, \mathrm{d}, m}$ is concave with respect to $\mathbf{P}_{m}$ \cite{boyd2004convex}, the second constraint is nonconvex. To sidestep this issue, we drop the second constraint and solve
\begin{alignat}{2}\label{p4}
&\underset{\{\mathbf{P}_{m}\}}{\mathrm{max}}\enspace&&\underset{k\in\llbracket K\rrbracket}{\mathrm{min}}\ \mathrm{SINR}_{\mathrm{MRT}, k}\bl\{\mathbf{P}_{m}\}_{\forall m}, \bm{\sigma}\br\notag\\
&\mathrm{subject}\ \mathrm{to}\enspace              &&P_{\mathrm{MRT}, \mathrm{d}, m}\bl\mathbf{P}_{m}, \sigma_{\mathrm{d}, m}\br\leq P_{\mathrm{d}},\enspace m\in\llbracket M\rrbracket,\notag\\
&                                                   &&\mathbf{P}_{m}\succeq\mathbf{0}_{K\times K},\enspace m\in\llbracket M\rrbracket.\tag{P4}
\end{alignat}

As a preliminary, we define $a_{m, k}=(1-\rho_{\mathrm{d}})\|\hat{\mathbf{g}}_{m, k}\|^{2}\geq 0$ in \eqref{vart0} so that the numerator of $\mathrm{SINR}_{\mathrm{MRT}, k}$ becomes
\begin{equation}\label{vart0_mrt}
\mathbb{E}\{|T_{0}|^{2}|\hat{\mathbf{g}}\}=\left(\sum_{m=1}^{M}a_{m, k}\sqrt{p_{m, k}}\right)^{2}
\end{equation}
where the absolute value is unnecessary unlike \eqref{vart0} because the sum of the terms inside the absolute value is nonnegative. Likewise,
\begin{align*}
&b_{m, i, k}=(1-\rho_{\mathrm{d}})\hat{\mathbf{g}}_{m, k}^{\mathrm{H}}\hat{\mathbf{g}}_{m, i}\in\mathbb{C},\\
&c_{m, i, k}=(1-\rho_{\mathrm{d}})^{2}\hat{\mathbf{g}}_{m, i}^{\mathrm{H}}\mathbf{C}_{\tilde{\mathbf{g}}_{m, k}}\hat{\mathbf{g}}_{m, i}+\rho_{\mathrm{d}}(1-\rho_{\mathrm{d}})\times\\
&\phantom{c_{m, i, k}}\mathrel{\phantom{=}}\mathrm{tr}\bl\bl\hat{\mathbf{g}}_{m, k}\hat{\mathbf{g}}_{m, k}^{\mathrm{H}}+\mathbf{C}_{\tilde{\mathbf{g}}_{m, k}}\br\mathrm{diag}\bl\hat{\mathbf{g}}_{m, i}\hat{\mathbf{g}}_{m, i}^{\mathrm{H}}\br\br\geq 0,\\
&d_{k}=\sum_{m=1}^{M}(1-\rho_{\mathrm{d}})\mathrm{tr}\bl\hat{\mathbf{g}}_{m, k}\hat{\mathbf{g}}_{m, k}^{\mathrm{H}}+\mathbf{C}_{\tilde{\mathbf{g}}_{m, k}}\br\sigma_{\mathrm{d}, m}^{2}\geq 0
\end{align*}
are introduced from \eqref{vart1}-\eqref{vart3} in a similar manner so that the denominator of $\mathrm{SINR}_{\mathrm{MRT}, k}$ is expressed as
\begin{align}\label{denominator}
&\mathbb{E}\{|T_{1}|^{2}|\hat{\mathbf{g}}\}+\mathbb{E}\{|T_{2}|^{2}|\hat{\mathbf{g}}\}+\mathbb{E}\{|T_{3}|^{2}|\hat{\mathbf{g}}\}+\sigma_{\mathrm{d}}^{2}=\notag\\
&\sum_{i\neq k}\left|\sum_{m=1}^{M}b_{m, i, k}\sqrt{p_{m, i}}\right|^{2}+\sum_{m=1}^{M}\sum_{i=1}^{K}c_{m, i, k}p_{m, i}+d_{k}+\sigma_{\mathrm{d}}^{2}.
\end{align}
Similarly, $P_{\mathrm{MRT}, \mathrm{d}, m}$ is linear with respect to $\mathbf{P}_{m}$, so $P_{\mathrm{MRT}, \mathrm{d}, m}$ admits the form
\begin{equation}\label{pdm_mrt}
P_{\mathrm{MRT}, \mathrm{d}, m}=\sum_{k=1}^{K}e_{m, k}p_{m, k}+f_{m}
\end{equation}
where
\begin{align*}
&e_{m, k}=(1-\rho_{\mathrm{d}})^{2}\|\mathbf{W}_{m}\hat{\mathbf{g}}_{m, k}\|^{2}+\rho_{\mathrm{d}}(1-\rho_{\mathrm{d}})\times\\
&\phantom{e_{m, k}}\mathrel{\phantom{=}}\mathrm{tr}\bl\mathbf{W}_{m}\mathrm{diag}\bl\hat{\mathbf{g}}_{m, k}\hat{\mathbf{g}}_{m, k}^{\mathrm{H}}\br\mathbf{W}_{m}^{\mathrm{H}}\br\geq 0,\\
&f_{m}=(1-\rho_{\mathrm{d}})\mathrm{tr}\bl\mathbf{W}_{m}\mathbf{W}_{m}^{\mathrm{H}}\br\sigma_{\mathrm{d}, m}^{2}\geq 0
\end{align*}
are given by \eqref{pdm_matrix_form}.

Now, let us introduce $x_{m, k}=\sqrt{p_{m, k}}$ in \eqref{vart0_mrt}-\eqref{pdm_mrt} so that \eqref{p4} is equivalent to
\begin{alignat}{2}\label{p5}
&\underset{\{x_{m, k}, y_{i, k}\}}{\mathrm{max}}\enspace&&\underset{k\in\llbracket K\rrbracket}{\mathrm{min}}\ \frac{\displaystyle\left(\sum_{m=1}^{M}a_{m, k}x_{m, k}\right)^{2}}{\displaystyle\sum_{i\neq k}y_{i, k}^{2}+\sum_{m=1}^{M}\sum_{i=1}^{K}c_{m, i, k}x_{m, i}^{2}+d_{k}+\sigma_{\mathrm{d}}^{2}}\notag\\
&\mathrm{subject}\ \mathrm{to}\enspace                  &&\sum_{k=1}^{K}e_{m, k}x_{m, k}^{2}+f_{m}\leq P_{\mathrm{d}},\enspace m\in\llbracket M\rrbracket,\notag\\
&                                                       &&\left|\sum_{m=1}^{M}b_{m, i, k}x_{m, i}\right|\leq y_{i, k},\enspace\{(i, k)\in\llbracket K\rrbracket^{2}|i\neq k\},\notag\\
&                                                       &&x_{m, k}\geq 0,\enspace(m, k)\in\llbracket M\rrbracket\times\llbracket K\rrbracket\tag{P5}
\end{alignat}
where $y_{i, k}$ is a dummy variable that upper bounds the left-hand side of the second constraint. The equivalence holds because the solution of \eqref{p5} satisfies the second inequality constraint with equality. To tackle \eqref{p5}, we present the following lemma. To simplify the notations in the lemma and sequel, we introduce the shorthand notations
\begin{align}
&\mathbf{x}_{k}=\begin{bmatrix}\sqrt{c_{1, 1, k}}x_{1, 1}&\cdots&\sqrt{c_{M, K, k}}x_{M, K}\end{bmatrix},\\
&\mathbf{y}_{k}=\begin{bmatrix}y_{1, k}&\cdots&y_{k-1, k}&y_{k+1, k}&\cdots&y_{K, k}\end{bmatrix}.
\end{align}

\begin{lemma}\label{lemma_1}
\eqref{p5} is a quasiconcave optimization problem.
\begin{proof}
To prove that \eqref{p5} is quasiconcave, let us analyze the objective function and constraints one by one. To show that the objective function is quasiconcave with respect to $\{x_{m, k}, y_{i, k}\}$, let us re-express the superlevel set
\begin{equation}
\underset{k\in\llbracket K\rrbracket}{\mathrm{min}}\ \frac{\displaystyle\left(\sum_{m=1}^{M}a_{m, k}x_{m, k}\right)^{2}}{\displaystyle\sum_{i\neq k}y_{i, k}^{2}+\sum_{m=1}^{M}\sum_{i=1}^{K}c_{m, i, k}x_{m, i}^{2}+d_{k}+\sigma_{\mathrm{d}}^{2}}\geq t
\end{equation}
as
\begin{align}
&\sqrt{\sum_{i\neq k}y_{i, k}^{2}+\sum_{m=1}^{M}\sum_{i=1}^{K}c_{m, i, k}x_{m, i}^{2}+d_{k}+\sigma_{\mathrm{d}}^{2}}=\notag\\
&\left\|\begin{bmatrix}\mathbf{x}_{k}&\mathbf{y}_{k}&\sqrt{d_{k}+\sigma_{\mathrm{d}}^{2}}\end{bmatrix}\right\|\leq\frac{1}{\sqrt{t}}\sum_{m=1}^{M}a_{m, k}x_{m, k},\enspace k\in\llbracket K\rrbracket.
\end{align}
Since second-order cone (SOC) constraints define convex sets \cite{boyd2004convex}, the superlevel set is convex, which proves that the objective function of \eqref{p5} is quasiconcave. Also, the first and second constraints of \eqref{p5} are quadratic and SOC constraints, which are convex \cite{boyd2004convex}. Therefore, \eqref{p5} is quasiconcave.
\end{proof}
\end{lemma}

According to Lemma \ref{lemma_1}, we can solve \eqref{p5} using the bisection method \cite{boyd2004convex}. The bisection method attains the global optimum of \eqref{p5} by solving the SOC programming (SOCP) feasibility problem
\begin{alignat}{2}\label{p6}
&\mathrm{find}\enspace                &&\{x_{m, k}, y_{i, k}\}\notag\\
&\mathrm{subject}\ \mathrm{to}\enspace&&\left\|\begin{bmatrix}\mathbf{x}_{k}&\mathbf{y}_{k}&\sqrt{d_{k}+\sigma_{\mathrm{d}}^{2}}\end{bmatrix}\right\|\leq\frac{1}{\sqrt{t}}\sum_{m=1}^{M}a_{m, k}x_{m, k},\notag\\
&                                     &&k\in\llbracket K\rrbracket,\notag\\
&                                     &&\sum_{k=1}^{K}e_{m, k}x_{m, k}^{2}+f_{m}\leq P_{\mathrm{d}},\enspace m\in\llbracket M\rrbracket,\notag\\
&                                     &&\left|\sum_{m=1}^{M}b_{m, i, k}x_{m, i}\right|\leq y_{i, k},\enspace\{(i, k)\in\llbracket K\rrbracket^{2}|i\neq k\},\notag\\
&                                     &&x_{m, k}\geq 0,\enspace(m, k)\in\llbracket M\rrbracket\times\llbracket K\rrbracket\tag{P6}
\end{alignat}
at each iteration where $t$ varies from iteration to iteration. To solve \eqref{p6}, we can use off-the-shelf convex solvers like CVX \cite{grant2014cvx}. Recall that the solution of \eqref{p5} corresponds to the solution of \eqref{p4}, which is equivalent to solving \eqref{p3} without the second constraint over $\{\mathbf{P}_{m}\}_{\forall m}$. After solving \eqref{p5}, we return to \eqref{p3} by $\{p_{m, k}\}_{\forall m, k}=\{x_{m, k}^{2}\}_{\forall m, k}$.

In the optimization over $\bm{\sigma}$, we drop the first constraint and solve \eqref{p3}. To proceed, note that $\mathrm{SINR}_{\mathrm{MRT}, k}$ is a nonincreasing function of $\bm{\sigma}$ as evident from \eqref{sinrk} and \eqref{vart3}. Also, $C_{\mathrm{MRT}, \mathrm{d}, m}$ is a nonincreasing function of $\sigma_{\mathrm{d}, m}$ according to \eqref{cdm_matrix_form}. Therefore, the global optimum of \eqref{p3} (without the first constraint) over $\bm{\sigma}$ is the one that equates the second inequality constraint. Since $C_{\mathrm{MRT}, \mathrm{d}, m}$ is differentiable with respect to $\sigma_{\mathrm{d}, m}$, the solution $\sigma_{\mathrm{d}, m}=C_{\mathrm{MRT}, \mathrm{d}, m}^{-1}\bl C_{\mathrm{d}}, \mathbf{P}_{m}\br$ of $C_{\mathrm{MRT}, \mathrm{d}, m}\bl\mathbf{P}_{m}, \sigma_{\mathrm{d}, m}\br=C_{\mathrm{d}}$ can be found numerically using root-finding algorithms like Newton's method.

After solving \eqref{p3} without the first constraint over $\bm{\sigma}$, it is possible that the first constraint of \eqref{p3} is violated. To make $\{\mathbf{P}_{m}\}_{\forall m}$ and $\bm{\sigma}$ feasible without violating the equality in the second inequality constraint, we project the current solution to the feasible set of \eqref{p3} by $p_{m, k}\coloneqq\mathrm{min}(P_{\mathrm{d}}/P_{\mathrm{MRT}, \mathrm{d}, m}, 1)p_{m, k}$ and $\sigma_{\mathrm{d}, m}\coloneqq\mathrm{min}(\sqrt{P_{\mathrm{d}}/P_{\mathrm{MRT}, \mathrm{d}, m}}, 1)\sigma_{\mathrm{d}, m}$, which concludes one iteration of the AO method. From \eqref{pdm_matrix_form}, \eqref{cdm_matrix_form}, and \eqref{p3}, it is apparent that such a scaling satisfies the first constraint, while retaining the equality in the second inequality constraint.

The AO method discussed until now is outlined in Algorithm \ref{algorithm_1}. In particular, Algorithm \ref{algorithm_1} attempts to solve \eqref{p3} as follows. The subproblem that Lines 6-15 solve corresponds to \eqref{p3} without the second constraint, which is equivalent to \eqref{p5}. Likewise, the subproblem that Line 17 solves is equivalent to \eqref{p3} without the first constraint. By leveraging the quasiconcavity of \eqref{p5} and monotonicity of $\mathrm{SINR}_{\mathrm{MRT}, k}$ and $C_{\mathrm{MRT}, \mathrm{d}, m}$, Algorithm \ref{algorithm_1} attains the global optima of the subproblems at each iteration.

\begin{algorithm}[t]
\caption{AO method for \eqref{p3}}\label{algorithm_1}
\begin{algorithmic}[1]
\State // $\{x_{m, k}(t, \bm{\sigma}), y_{i, k}(t, \bm{\sigma})\}$ is the solution of \eqref{p6}
\State Set $t_{\mathrm{min}}$, $t_{\mathrm{max}}$, and $\epsilon$ for the bisection method
\State Initialize $\bm{\sigma}$
\While {termination condition}
\State // optimization over $\{\mathbf{P}_{m}\}_{\forall m}$
\State $t_{\mathrm{lo}}\coloneqq t_{\mathrm{min}}$ and $t_{\mathrm{up}}\coloneqq t_{\mathrm{max}}$
\While {$t_{\mathrm{up}}-t_{\mathrm{lo}}>\epsilon$}
\State $t\coloneqq(t_{\mathrm{lo}}+t_{\mathrm{up}})/2$ and solve \eqref{p6}
\If {\eqref{p6} is feasible}
\State $t_{\mathrm{lo}}\coloneqq t$
\Else
\State $t_{\mathrm{up}}\coloneqq t$
\EndIf
\EndWhile
\State $p_{m, k}\coloneqq x_{m, k}^{2}(t_{\mathrm{lo}}, \bm{\sigma}),\enspace(m, k)\in\llbracket M\rrbracket\times\llbracket K\rrbracket$
\State // optimization over $\bm{\sigma}$
\State $\sigma_{\mathrm{d}, m}\coloneqq C_{\mathrm{MRT}, \mathrm{d}, m}^{-1}\bl C_{\mathrm{d}}, \mathbf{P}_{m}\br,\enspace m\in\llbracket M\rrbracket$
\State // projection to the feasible set of \eqref{p3}
\State $\kappa_{m}\coloneqq P_{\mathrm{d}}/P_{\mathrm{MRT}, \mathrm{d}, m}\bl\mathbf{P}_{m}, \sigma_{\mathrm{d}, m}\br,\enspace m\in\llbracket M\rrbracket$
\State $p_{m, k}\coloneqq\mathrm{min}(\kappa_{m}, 1)p_{m, k},\enspace(m, k)\in\llbracket M\rrbracket\times\llbracket K\rrbracket$
\State $\sigma_{\mathrm{d}, m}\coloneqq\mathrm{min}(\sqrt{\kappa_{m}}, 1)\sigma_{\mathrm{d}, m},\enspace m\in\llbracket M\rrbracket$
\EndWhile
\end{algorithmic}
\end{algorithm}

The caveat, however, is that the subproblems are not strictly the subproblems of \eqref{p3} in the sense that the first and second constraints are missing. Therefore, there is no guarantee for the objective function of \eqref{p3} evaluated at $\{p_{m, k}, \sigma_{\mathrm{d}, m}\}$ in Lines 20 and 21 to increase at each iteration, which implies that Algorithm \ref{algorithm_1} can diverge, at least theoretically. This is the main shortcoming of the heuristic approach proposed in \cite{8678745} adopted to our case that sidesteps the nonconvexity of the original problem by discarding some of the constraints. The simulation results in Section \ref{section_4}, nevertheless, demonstrate that Algorithm \ref{algorithm_1} performs well in practice, and the discussion until now is not as problematic. The complexity of Algorithm \ref{algorithm_1} is high because Line 8 solves the SOCP feasibility problem at each iteration, which is prohibitive for large systems like cell-free massive MIMO.

\subsection{Max-Min Fairness Algorithm for the ZF Precoder}
In this subsection, the max-min fairness problem for the ZF precoder is addressed. The distinct feature of the proposed novel algorithm is that the convergence is guaranteed, which is in contrast to the case of the MRT precoder. Also, the complexity of the proposed algorithm is low because there is no convex optimization problem involved.

We adopt the ZF precoder from \cite{7917284, 9420030} that eliminates the inter-user interference by setting the linear precoder as the pseudoinverse of the channel estimate, namely\footnote{The right inverse of the channel estimate exists if $K\leq MR$, and the right inverse that eliminates the inter-user interference is given by the pseudoinverse of the channel estimate.}
\begin{equation}\label{pinv}
\begin{bmatrix}\mathbf{f}_{1, 1}&\cdots&\mathbf{f}_{1, K}\\\vdots&\ddots&\vdots\\\mathbf{f}_{M, 1}&\cdots&\mathbf{f}_{M, K}\end{bmatrix}=\begin{bmatrix}\hat{\mathbf{g}}_{1, 1}^{\mathrm{H}}&\cdots&\hat{\mathbf{g}}_{M, 1}^{\mathrm{H}}\\\vdots&\ddots&\vdots\\\hat{\mathbf{g}}_{1, K}^{\mathrm{H}}&\cdots&\hat{\mathbf{g}}_{M, K}^{\mathrm{H}}\end{bmatrix}^{\dagger},
\end{equation}
so that
\begin{equation}
\sum_{m=1}^{M}\hat{\mathbf{g}}_{m, k}^{\mathrm{H}}\mathbf{f}_{m, i}=\begin{cases}1\ \text{if}\ i=k\\0\ \text{if}\ i\neq k\end{cases},
\end{equation}
while constraining $\{p_{m, k}\}_{\forall m, k}$ to depend only on $k$. Hence, we drop the subscript $m$ from $\{p_{m, k}\}_{\forall m, k}$ and introduce the shorthand notation
\begin{equation}
\mathbf{P}=\mathrm{blockdiag}(p_{1}, \dots, p_{K}).
\end{equation}
As a result, \eqref{sinrk} depends not on $\{\mathbf{P}_{m}\}_{\forall m}$ but $\mathbf{P}$ as
\begin{align}\label{sinrk_zf}
&\mathrm{SINR}_{\mathrm{ZF}, k}\bl\mathbf{P}, \bm{\sigma}\br=\notag\\
&\frac{\mathbb{E}\{|T_{0}|^{2}|\hat{\mathbf{g}}\}}{\mathbb{E}\{|T_{1}|^{2}|\hat{\mathbf{g}}\}+\mathbb{E}\{|T_{2}|^{2}|\hat{\mathbf{g}}\}+\mathbb{E}\{|T_{3}|^{2}|\hat{\mathbf{g}}\}+\sigma_{\mathrm{d}}^{2}},
\end{align}
while \eqref{vart0}-\eqref{vart3} become
\begin{align}
&\mathbb{E}\{|T_{0}|^{2}|\hat{\mathbf{g}}\}=(1-\rho_{\mathrm{d}})^{2}p_{k},\label{vart0_zf}\\
&\mathbb{E}\{|T_{1}|^{2}|\hat{\mathbf{g}}\}=\sum_{m=1}^{M}(1-\rho_{\mathrm{d}})^{2}\mathbf{f}_{m, k}^{\mathrm{H}}\mathbf{C}_{\tilde{\mathbf{g}}_{m, k}}\mathbf{f}_{m, k}p_{k},\\
&\mathbb{E}\{|T_{2}|^{2}|\hat{\mathbf{g}}\}=\sum_{m=1}^{M}\sum_{i\neq k}(1-\rho_{\mathrm{d}})^{2}\mathbf{f}_{m, i}^{\mathrm{H}}\mathbf{C}_{\tilde{\mathbf{g}}_{m, k}}\mathbf{f}_{m, i}p_{i},\\
&\mathbb{E}\{|T_{3}|^{2}|\hat{\mathbf{g}}\}=(1-\rho_{\mathrm{d}})\sum_{m=1}^{M}\notag\\
&\mathrm{tr}\bl\bl\hat{\mathbf{g}}_{m, k}\hat{\mathbf{g}}_{m, k}^{\mathrm{H}}+\mathbf{C}_{\tilde{\mathbf{g}}_{m, k}}\br\bl\rho_{\mathrm{d}}\mathrm{diag}\bl\mathbf{F}_{m}\mathbf{P}\mathbf{F}_{m}^{\mathrm{H}}\br+\sigma_{\mathrm{d}, m}^{2}\mathbf{I}_{R}\br\br\label{vart3_zf}
\end{align}
where \eqref{vart0_zf}-\eqref{vart3_zf} constitute the numerator and denominator of $\mathrm{SINR}_{\mathrm{ZF}, k}$. Also, \eqref{pdm_matrix_form} and \eqref{cdm_matrix_form} depend not on $\mathbf{P}_{m}$ but $\mathbf{P}$ as
\begin{align}
&P_{\mathrm{ZF}, \mathrm{d}, m}\bl\mathbf{P}, \sigma_{\mathrm{d}, m}\br=\notag\\
&\quad(1-\rho_{\mathrm{d}})^{2}\mathrm{tr}\bl\mathbf{W}_{m}\mathbf{F}_{m}\mathbf{P}\mathbf{F}_{m}^{\mathrm{H}}\mathbf{W}_{m}^{\mathrm{H}}\br+\notag\\
&\quad\rho_{\mathrm{d}}(1-\rho_{\mathrm{d}})\mathrm{tr}\bl\mathbf{W}_{m}\mathrm{diag}\bl\mathbf{F}_{m}\mathbf{P}\mathbf{F}_{m}^{\mathrm{H}}\br\mathbf{W}_{m}^{\mathrm{H}}\br+\notag\\
&\quad(1-\rho_{\mathrm{d}})\mathrm{tr}\bl\mathbf{W}_{m}\mathbf{W}_{m}^{\mathrm{H}}\br\sigma_{\mathrm{d}, m}^{2},\label{pdm_zf}\\
&C_{\mathrm{ZF}, \mathrm{d}, m}\bl\mathbf{P}, \sigma_{\mathrm{d}, m}\br=\log_{2}\mathrm{det}\left(\mathbf{I}_{R}+\frac{1}{\sigma_{\mathrm{d}, m}^{2}}\mathbf{F}_{m}\mathbf{P}\mathbf{F}_{m}^{\mathrm{H}}\right).\label{cdm_zf}
\end{align}

In this subsection, the goal is to solve the max-min fairness problem \cite{7827017, 7917284}
\begin{align}\label{p7}
&\underset{\mathbf{P}\succeq\mathbf{0}_{K\times K}, \bm{\sigma}\succeq\mathbf{0}_{M\times M}}{\mathrm{max}}\enspace&&\underset{k\in\llbracket K\rrbracket}{\mathrm{min}}\ \mathrm{SINR}_{\mathrm{ZF}, k}\bl\mathbf{P}, \bm{\sigma}\br\notag\\
&\mathrm{subject}\ \mathrm{to}\enspace                                                                             &&P_{\mathrm{ZF}, \mathrm{d}, m}\bl\mathbf{P}, \sigma_{\mathrm{d}, m}\br\leq P_{\mathrm{d}},\enspace m\in\llbracket M\rrbracket,\notag\\
&                                                                                                                  &&C_{\mathrm{ZF}, \mathrm{d}, m}\bl\mathbf{P}, \sigma_{\mathrm{d}, m}\br\leq C_{\mathrm{d}},\enspace m\in\llbracket M\rrbracket.\tag{P7}
\end{align}
In general, the joint optimization over $\mathbf{P}$ and $\bm{\sigma}$ is a daunting task because \eqref{p7} is nonconvex. Hence, we develop an AO method that alternates between $\mathbf{P}$ and $\bm{\sigma}$. In contrast to Algorithm \ref{algorithm_1}, the AO method for \eqref{p7} takes into account all the constraints at each subproblem. Furthermore, since the global optima of the subproblems are attained at each iteration, the convergence to the local optimum of \eqref{p7} is guaranteed as no constraint is discarded.

As a preliminary, let us establish some observations on how the objective and constraint functions of \eqref{p7} depend on $\mathbf{P}$ and $\bm{\sigma}$ by inspection:
\begin{itemize}
\item $\mathrm{SINR}_{\mathrm{ZF}, k}$ is a quasilinear function of $\mathbf{P}$.
\item $P_{\mathrm{ZF}, \mathrm{d}, m}$ is a linear function of $\mathbf{P}$.
\item $C_{\mathrm{ZF}, \mathrm{d}, m}$ is a concave function of $\mathbf{P}$.
\item $\mathrm{SINR}_{\mathrm{ZF}, k}$ is a nonincreasing function of $\bm{\sigma}$.
\item $P_{\mathrm{ZF}, \mathrm{d}, m}$ is a nondecreasing function of $\sigma_{\mathrm{d}, m}$.
\item $C_{\mathrm{ZF}, \mathrm{d}, m}$ is a nonincreasing function of $\sigma_{\mathrm{d}, m}$.
\end{itemize}
To explain why the first observation holds, observe that the numerator and denominator of $\mathrm{SINR}_{\mathrm{ZF}, k}$ are linear with respect to $\mathbf{P}$ as evident from \eqref{sinrk_zf}-\eqref{vart3_zf}. Since the sublevel and superlevel sets of such a function are convex \cite{boyd2004convex}, the first observation is valid. The remaining observations follow from \eqref{sinrk_zf}-\eqref{cdm_zf}.

Now, let us address the optimization over $\mathbf{P}$. To solve \eqref{p7} for fixed $\bm{\sigma}$, we adopt the bisection method \cite{boyd2004convex} inspired by the approach proposed in \cite{7917284}. In particular, the bisection method attains the global optimum of \eqref{p7} over $\mathbf{P}$ by solving the feasibility problem
\begin{align}\label{p8}
&\mathrm{find}\enspace                &&\mathbf{P}\succeq\mathbf{0}_{K\times K}\notag\\
&\mathrm{subject}\ \mathrm{to}\enspace&&\mathrm{SINR}_{\mathrm{ZF}, k}\bl\mathbf{P}, \bm{\sigma}\br\geq t,\enspace k\in\llbracket K\rrbracket,\notag\\
&                                     &&P_{\mathrm{ZF}, \mathrm{d}, m}\bl\mathbf{P}, \sigma_{\mathrm{d}, m}\br\leq P_{\mathrm{d}},\enspace m\in\llbracket M\rrbracket,\notag\\
&                                     &&C_{\mathrm{ZF}, \mathrm{d}, m}\bl\mathbf{P}, \sigma_{\mathrm{d}, m}\br\leq C_{\mathrm{d}},\enspace m\in\llbracket M\rrbracket\tag{P8}
\end{align}
at each iteration where $t$ evolves from iteration to iteration. Unlike \cite{7917284} where each iteration of the bisection method is cast as a convex feasibility problem, however, \eqref{p8} is nonconvex. To see why, recall that $C_{\mathrm{ZF}, \mathrm{d}, m}$ is concave with respect to $\mathbf{P}$ \cite{boyd2004convex}, whose sublevel set is nonconvex. Therefore, the bisection method approach \cite{7917284} that leverages the convexity of the feasibility problem is not applicable to our case.

To address \eqref{p8}, we introduce a lemma from \cite{7917284} to prove a statement that answers how to solve \eqref{p8}.
\begin{lemma}\label{lemma_2}
Consider diagonal matrices $\mathbf{P}$ and $\mathbf{P}'\succ\mathbf{0}_{K\times K}$ that satisfy $\mathrm{SINR}_{\mathrm{ZF}, k}\bl\mathbf{P}, \bm{\sigma}\br=t$ and $\mathrm{SINR}_{\mathrm{ZF}, k}\bl\mathbf{P}', \bm{\sigma}\br\geq t$ for $k\in\llbracket K\rrbracket$. Then, $\mathbf{P}'\succeq\mathbf{P}\succeq\mathbf{0}_{K\times K}$. \cite{7917284}
\end{lemma}
Lemma \ref{lemma_2} holds as long as the linear coefficients of $\mathbf{P}$ in the numerator and denominator of $\mathrm{SINR}_{\mathrm{ZF}, k}$ are nonnegative. Since the condition is true as evident from \eqref{sinrk_zf}-\eqref{vart3_zf}, we adopt Lemma \ref{lemma_2} to establish the following lemma that answers how to address \eqref{p8}.
\begin{lemma}\label{lemma_3}
\eqref{p8} is feasible if and only if the solution $\mathbf{P}$ of $\mathrm{SINR}_{\mathrm{ZF}, k}\bl\mathbf{P}, \bm{\sigma}\br=t$ for $k\in\llbracket K\rrbracket$ satisfies the constraints of \eqref{p8}.
\begin{proof}
Since ``if" is trivial, let us prove ``only if." Assume that \eqref{p8} is feasible. Then, there exists $\mathbf{P}'\succ\mathbf{0}_{K\times K}$ that satisfies the constraints of \eqref{p8}. Since Lemma \ref{lemma_2} prescribes the solution $\mathbf{P}$ of $\mathrm{SINR}_{\mathrm{ZF}, k}\bl\mathbf{P}, \bm{\sigma}\br=t$ for $k\in\llbracket K\rrbracket$ to satisfy $\mathbf{P}'\succeq\mathbf{P}\succeq\mathbf{0}_{K\times K}$, $\mathbf{P}$ satisfies the nonnegative definite constraint. Also, $\mathbf{P}$ trivially satisfies the first inequality constraint with equality. To check the second constraint, observe that the linear coefficients of $\mathbf{P}$ in $P_{\mathrm{ZF}, \mathrm{d}, m}$ are nonnegative as evident from \eqref{pdm_zf}. Since $\mathbf{P}'\succeq\mathbf{P}$, we conclude that $P_{\mathrm{ZF}, \mathrm{d}, m}\bl\mathbf{P}, \sigma_{\mathrm{d}, m}\br\leq P_{\mathrm{ZF}, \mathrm{d}, m}\bl\mathbf{P}', \sigma_{\mathrm{d}, m}\br\leq P_{\mathrm{d}}$ for $m\in\llbracket M\rrbracket$, which implies that $\mathbf{P}$ satisfies the second constraint. To verify the third constraint, consider the expansion
\begin{align}
&\underbrace{\mathbf{I}_{R}+\frac{1}{\sigma_{\mathrm{d}, m}^{2}}\mathbf{F}_{m}\mathbf{P}'\mathbf{F}_{m}^{\mathrm{H}}}_{=\mathbf{A}+\mathbf{B}}=\notag\\
&\underbrace{\frac{1}{\sigma_{\mathrm{d}, m}^{2}}\mathbf{F}_{m}\bl\mathbf{P}'-\mathbf{P}\br\mathbf{F}_{m}^{\mathrm{H}}}_{=\mathbf{A}\succeq\mathbf{0}_{K\times K}}+\underbrace{\mathbf{I}_{R}+\frac{1}{\sigma_{\mathrm{d}, m}^{2}}\mathbf{F}_{m}\mathbf{P}\mathbf{F}_{m}^{\mathrm{H}}}_{=\mathbf{B}\succeq\mathbf{0}_{K\times K}}
\end{align}
where the nonnegative definiteness of $\mathbf{A}$ and $\mathbf{B}$ follows from $\mathbf{P}'-\mathbf{P}\succeq\mathbf{0}_{K\times K}$ and $\mathbf{P}\succeq\mathbf{0}_{K\times K}$. Since Minkowski's inequality \cite{104312} yields $\mathrm{det}\bl\mathbf{A}+\mathbf{B}\br\geq\mathrm{det}\bl\mathbf{A}\br+\mathrm{det}\bl\mathbf{B}\br\geq\mathrm{det}\bl\mathbf{B}\br$, we conclude that $C_{\mathrm{ZF}, \mathrm{d}, m}\bl\mathbf{P}, \sigma_{\mathrm{d}, m}\br\leq C_{\mathrm{ZF}, \mathrm{d}, m}\bl\mathbf{P}', \sigma_{\mathrm{d}, m}\br\leq C_{\mathrm{d}}$ for $m\in\llbracket M\rrbracket$ by the monotonicity of logarithm. Therefore, $\mathbf{P}$ satisfies the third constraint.
\end{proof}
\end{lemma}

Lemma \ref{lemma_3} provides a guideline for checking the feasibility of \eqref{p8}. First, we find the solution $\mathbf{P}$ of
\begin{equation}\label{linear_equation}
\mathrm{SINR}_{\mathrm{ZF}, k}\bl\mathbf{P}, \bm{\sigma}\br=t,\enspace k\in\llbracket K\rrbracket.
\end{equation}
Since the numerator and denominator of $\mathrm{SINR}_{\mathrm{ZF}, k}$ are linear with respect to $\mathbf{P}$, \eqref{linear_equation} is equivalent to a system of $K$ linear equations with $K$ unknowns with a unique solution. Therefore, \eqref{p8} is feasible if and only if $\mathbf{P}$ is nonnegative definite and satisfies the second and third constraints of \eqref{p8} but infeasible otherwise.

\begin{algorithm}[t]
\caption{AO method for \eqref{p7}}\label{algorithm_2}
\begin{algorithmic}[1]
\State // $\mathbf{P}(t, \bm{\sigma})$ is the solution of \eqref{p8}
\State Set $t_{\mathrm{min}}$, $t_{\mathrm{max}}$, and $\epsilon$ for the bisection method
\State Initialize $\bm{\sigma}$
\While {termination condition}
\State // optimization over $\mathbf{P}$
\State $t_{\mathrm{lo}}\coloneqq t_{\mathrm{min}}$ and $t_{\mathrm{up}}\coloneqq t_{\mathrm{max}}$
\While {$t_{\mathrm{up}}-t_{\mathrm{lo}}>\epsilon$}
\State $t\coloneqq(t_{\mathrm{lo}}+t_{\mathrm{up}})/2$ and solve \eqref{p8}
\If {\eqref{p8} is feasible}
\State $t_{\mathrm{lo}}\coloneqq t$
\Else
\State $t_{\mathrm{up}}\coloneqq t$
\EndIf
\EndWhile
\State $\mathbf{P}\coloneqq\mathbf{P}(t_{\mathrm{lo}}, \bm{\sigma})$
\State // optimization over $\bm{\sigma}$
\State $\sigma_{\mathrm{d}, m}\coloneqq C_{\mathrm{ZF}, \mathrm{d}, m}^{-1}\bl C_{\mathrm{d}}, \mathbf{P}\br,\enspace m\in\llbracket M\rrbracket$
\EndWhile
\end{algorithmic}
\end{algorithm}

Moving on to the optimization over $\bm{\sigma}$ where \eqref{p7} is solved for fixed $\mathbf{P}$, note that $\mathrm{SINR}_{\mathrm{ZF}, k}$ is a nonincreasing function of $\bm{\sigma}$. Therefore, the objective function increases as $\bm{\sigma}$ decreases. Meanwhile, the first constraint becomes more feasible as $\bm{\sigma}$ decreases because $P_{\mathrm{ZF}, \mathrm{d}, m}$ is a nondecreasing function of $\sigma_{\mathrm{d}, m}$. Since $C_{\mathrm{ZF}, \mathrm{d}, m}$ is a nonincreasing function of $\sigma_{\mathrm{d}, m}$, however, we observe that the global optimum of \eqref{p7} over $\bm{\sigma}$ is the one that satisfies the second inequality constraint with equality. By taking advantage of the differentiability of $C_{\mathrm{ZF}, \mathrm{d}, m}$ with respect to $\sigma_{\mathrm{d}, m}$, we can find the solution $\sigma_{\mathrm{d}, m}=C_{\mathrm{ZF}, \mathrm{d}, m}^{-1}\bl C_{\mathrm{d}}, \mathbf{P}\br$ of $C_{\mathrm{ZF}, \mathrm{d}, m}\bl\mathbf{P}, \sigma_{\mathrm{d}, m}\br=C_{\mathrm{d}}$ using Newton's method.

The AO method discussed until now is presented in Algorithm \ref{algorithm_2}. In essence, Lines 6-15 and 17 solve \eqref{p7} over $\mathbf{P}$ and $\bm{\sigma}$ in an alternating fashion. The distinct feature of the AO method for \eqref{p7} is that Algorithm \ref{algorithm_2} takes into account all the constraints at each subproblem. Since Algorithm \ref{algorithm_2} attains the global optima of the subproblems at each iteration by utilizing Lemma \ref{lemma_3} and the monotonicity of $\mathrm{SINR}_{\mathrm{ZF}, k}$, $P_{\mathrm{ZF}, \mathrm{d}, m}$, and $C_{\mathrm{ZF}, \mathrm{d}, m}$, Algorithm \ref{algorithm_2} converges to the local optimum of \eqref{p7}. Furthermore, the complexity of Algorithm \ref{algorithm_2} is low because each iteration of the bisection method in Line 8 solves a system of linear equations, which is computationally more efficient than solving a convex optimization problem at each iteration like Algorithm \ref{algorithm_1}.

\section{Simulation Results}\label{section_4}
In this section, we evaluate energy efficiency, spectral efficiency, and channel estimation error of cell-free mmWave massive MIMO systems with low-capacity fronthaul links and low-resolution ADC/DACs. The system under consideration operates as follows. The control unit performs channel estimation based on the proposed scheme in Section \ref{section_2a}. Then, the control unit uses the channel estimate to precode the data signal in the downlink data transmission phase. The MRT and ZF precoders are adopted, whose power coefficient $\{\mathbf{P}_{m}\}_{\forall m}$ and codebook parameter $\bm{\sigma}$ are optimized by Algorithms \ref{algorithm_1} and \ref{algorithm_2}. The base stations that act as radio units use the long-term CSI of the $K/M$ nearest users to configure the RF combiners. In particular, the $R$ columns of $\mathbf{W}_{m}$ are set as the top-$R$ eigenvectors of $\mathbf{C}_{\mathbf{h}_{m, k_{m, 1}}}+\cdots+\mathbf{C}_{\mathbf{h}_{m, k_{m, K/M}}}$ where $\{k_{m, 1}, \dots, k_{m, K/M}\}$ is the index set of the $K/M$ users nearest to the $m$-th base station \cite{7919262}. Since the resulting RF combiners are semi-unitary and not necessarily unit-modulus, we use the alternating projection method \cite{1377501, 8333733} to project the RF combiners to the unit-modulus space. In particular, we repeat projecting the semi-unitary RF combiners to the unit-modulus space and projecting the unit-modulus RF combiners back to the semi-unitary space until convergence. The RF precoders in the downlink are identical to the RF combiners in the uplink. The training signals of the users are chosen from the columns of the discrete Fourier transform (DFT) matrix of size $T$. Throughout the simulations, we assume that the uplink-downlink fronthaul capacity and ADC/DAC resolution are symmetric as $C=C_{\mathrm{u}}=C_{\mathrm{d}}$ and $B=B_{\mathrm{u}}=B_{\mathrm{d}}$.

To evaluate the performance of Algorithms \ref{algorithm_1} and \ref{algorithm_2}, we adopt the heuristic approach of \cite{8678745} for the ZF precoder to solve \eqref{p7} as a baseline. In addition, quantizer-based cell-free massive MIMO \cite{8756265, 9123382} that restricts the fronthaul capacity and ADC/DAC resolution to be the same is adopted as another baseline. In particular, quantizer-based cell-free massive MIMO assumes $C/(2R)=B$ so that the ADC/DAC quantization is sufficient for the signal to be conveyed through fronthaul links, and there is no need for the fronthaul compression to be performed. For quantizer-based cell-free massive MIMO, the power allocation schemes that achieve the max-min fairness for the MRT and ZF precoders proposed in \cite{8756265, 9123382} are adopted.

The simulation parameters are set according to the Dense Urban-eMBB scenario in ITU-R M.2412-0 \cite{itu2017guidelines}. In particular, the carrier frequency, bandwidth, base station transmit power, user transmit power, base station noise figure, and user noise figure under consideration are $f_{c}=30$ GHz, $W=80$ MHz, $P_{\mathrm{d}}=33$ dBm, $P_{\mathrm{u}}=23$ dBm, $\mathrm{NF}_{\mathrm{u}}=7$ dB, and $\mathrm{NF}_{\mathrm{d}}=10$ dB. Hence, the noise power is $\sigma_{\mathrm{u}}^{2}=W\cdot N_{0}\cdot\mathrm{NF}_{\mathrm{u}}$ and $\sigma_{\mathrm{d}}^{2}=W\cdot N_{0}\cdot\mathrm{NF}_{\mathrm{d}}$ where $N_{0}=-174$ dBm/Hz is the noise spectral density. Also, the channel in \eqref{channel} is generated as specified in ITU-R M.2412 with the uniform planar array (UPA) geometry. The simulation setup is $M=16$, $K=32$, $N=16$, $R=4$, $T=K$, $W_{c}=180$ kHz, and $T_{c}=10$ ms where $W_{c}$ and $T_{c}$ denote the coherence bandwidth and coherence time. The base stations and users are randomly located in a square area of $250\times 250$ m$^{2}$, which is wrapped around to avoid the boundary effect.

To evaluate energy efficiency of various schemes, let us introduce the power consumption model under consideration. In particular, we focus on the power dissipated by the base stations and fronthaul links. The power consumed by one $B$-bit ADC/DAC is modeled as $P_{\mathrm{ADC}}=\mathrm{FOM}\cdot F_{s}\cdot 2^{B}$ \cite{4403893} where $\mathrm{FOM}$ and $F_{s}$ represent the figure of merit and sampling frequency. The power consumption of one RF chain is $P_{\mathrm{RF}}=2P_{\mathrm{LPF}}+2P_{\mathrm{M}}+P_{\mathrm{PS}}$ where $P_{\mathrm{LPF}}$, $P_{\mathrm{M}}$, and $P_{\mathrm{PS}}$ denote the power consumption of one low-pass filter, mixer, and 90$^{\circ}$ phase shifter. The power that the power amplifiers at the $m$-th base station consume in one coherence period is modeled as $P_{\mathrm{PA}, m}=(W_{c}T_{c}-T)/(W_{c}T_{c})\cdot\mathbb{E}\{\|\mathbf{x}_{\mathrm{d}, m}\|^{2}\}/\eta$ \cite{8333733} where $\eta$ is the power-added efficiency (PAE). Then, the power consumption of the $m$-th base station is $P_{\mathrm{BS}, m}=P_{\mathrm{PA}, m}+P_{\mathrm{LO}}+R(2P_{\mathrm{ADC}}+P_{\mathrm{RF}})$ where $P_{\mathrm{LO}}$ is the power consumption of one local oscillator. Furthermore, by incorporating the power consumed by the fronthaul links of $C$ bps/Hz, the minimum energy efficiency lower bound is defined as
\begin{equation}
\mathrm{EE}=\frac{\displaystyle\frac{W_{c}T_{c}-T}{W_{c}T_{c}}\mathbb{E}\left\{\underset{k\in\llbracket K\rrbracket}{\mathrm{min}}\ W\log_{2}(1+\mathrm{SINR}_{k})\right\}}{\displaystyle \sum_{m=1}^{M}P_{\mathrm{BS}, m}+M\cdot W\cdot C\cdot P_{\mathrm{FH}}}
\end{equation}
where $P_{\mathrm{FH}}$ is the traffic-dependent power consumption of one fronthaul link \cite{8891922}, while the expectation is over the coherence block, $\hat{\mathbf{g}}$ in particular. To provide a sense of how future wireless communications are expected to perform, the power consumption parameters are configured in an optimistic fashion such that the most efficient hardware implementation reported in the literature is adopted as shown in Table \ref{table_1}.

\begin{table}[t]
\centering
\caption{Power consumption parameters}\label{table_1}
\begin{tabular}{|c|c|c|c|}
\hline
\textbf{Parameter}&\textbf{Value}&\textbf{Parameter}&\textbf{Value}\\
\hline
$\mathrm{FOM}$ \cite{murmann2021adc}&1432.1 fJ/conversion-step&$P_{\mathrm{LO}}$ \cite{4684642}&22.5 mW\\
\hline
$F_{s}$ \cite{8333733}&1 GHz&$P_{\mathrm{LPF}}$ \cite{rangan2013energy}&14 mW\\
\hline
$P_{\mathrm{FH}}$ \cite{8891922}&2 W/Gbps&$P_{\mathrm{M}}$ \cite{5940650}&0.3 mW\\
\hline
$\eta$ \cite{8692752}&46\%&$P_{\mathrm{PS}}$ \cite{marcu2011cristian}&3 mW\\
\hline
\end{tabular}
\end{table}

\begin{figure}[t]
\centering
\includegraphics[width=1\columnwidth]{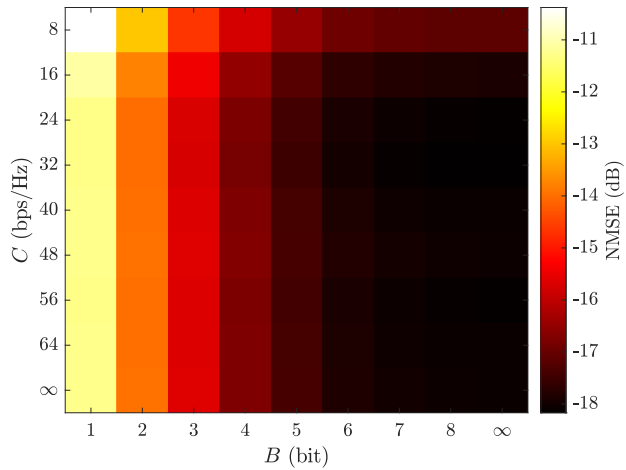}
\caption[caption]{The NMSE of the channel estimation error for $C=8, 16, 24, \dots, 64, \infty$ bps/Hz and $B=1, \dots, 8, \infty$ bits.}\label{figure_2}
\end{figure}

In the first simulation, the channel estimation error is investigated for $C=8, 16, 24, \dots, 64, \infty$ bps/Hz and $B=1, \dots, 8, \infty$ bits where the normalized MSE (NMSE) of the channel estimation error is defined as
\begin{equation}
\mathrm{NMSE}=\frac{\displaystyle\sum_{m=1}^{M}\mathbb{E}\{\|\mathbf{h}_{m}-\hat{\mathbf{h}}_{m}\|^{2}\}}{\displaystyle\sum_{m=1}^{M}\mathbb{E}\{\|\mathbf{h}_{m}\|^{2}\}}.
\end{equation}
According to Fig. \ref{figure_2}, observe that $C\geq 16$ bps/Hz, which is virtually equivalent to $C/(2R)\geq 2$ bps/Hz per IQ components, is sufficient to approach the NMSE of infinite-capacity fronthaul links. The low NMSE is attained by the codebook with the i.i.d. compression noise optimized so as to minimize the channel estimation error as proposed in Section \ref{section_2a}. The implication of such a phenomenon is that the gain obtained by using the codebook with the colored compression noise at the expense of the increased complexity is not as rewarding. This is the reason why we abandoned the codebook with the colored compression noise.

\begin{figure}[t]
\centering
\includegraphics[width=1\columnwidth]{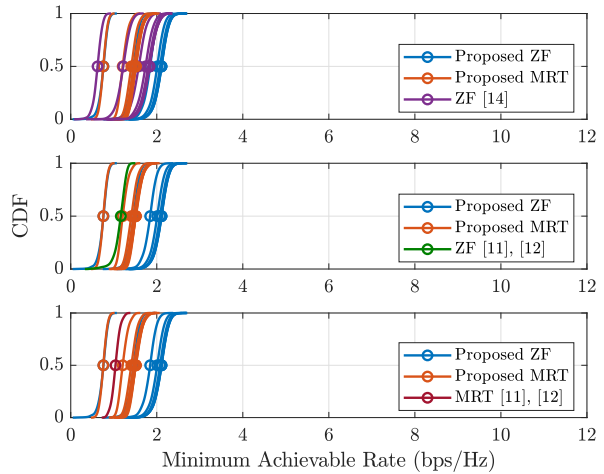}
\caption[caption]{The CDF of the minimum achievable rate lower bound of the users for $C=8$ bps/Hz. In each subplot, the lines with the same color correspond to $B=1, \dots, 8$ bits from left to right. The MRT and ZF precoders proposed in \cite{8756265, 9123382} are constrained to $C/(2R)=B=1$.}\label{figure_3}
\end{figure}

\begin{figure}[t]
\centering
\includegraphics[width=1\columnwidth]{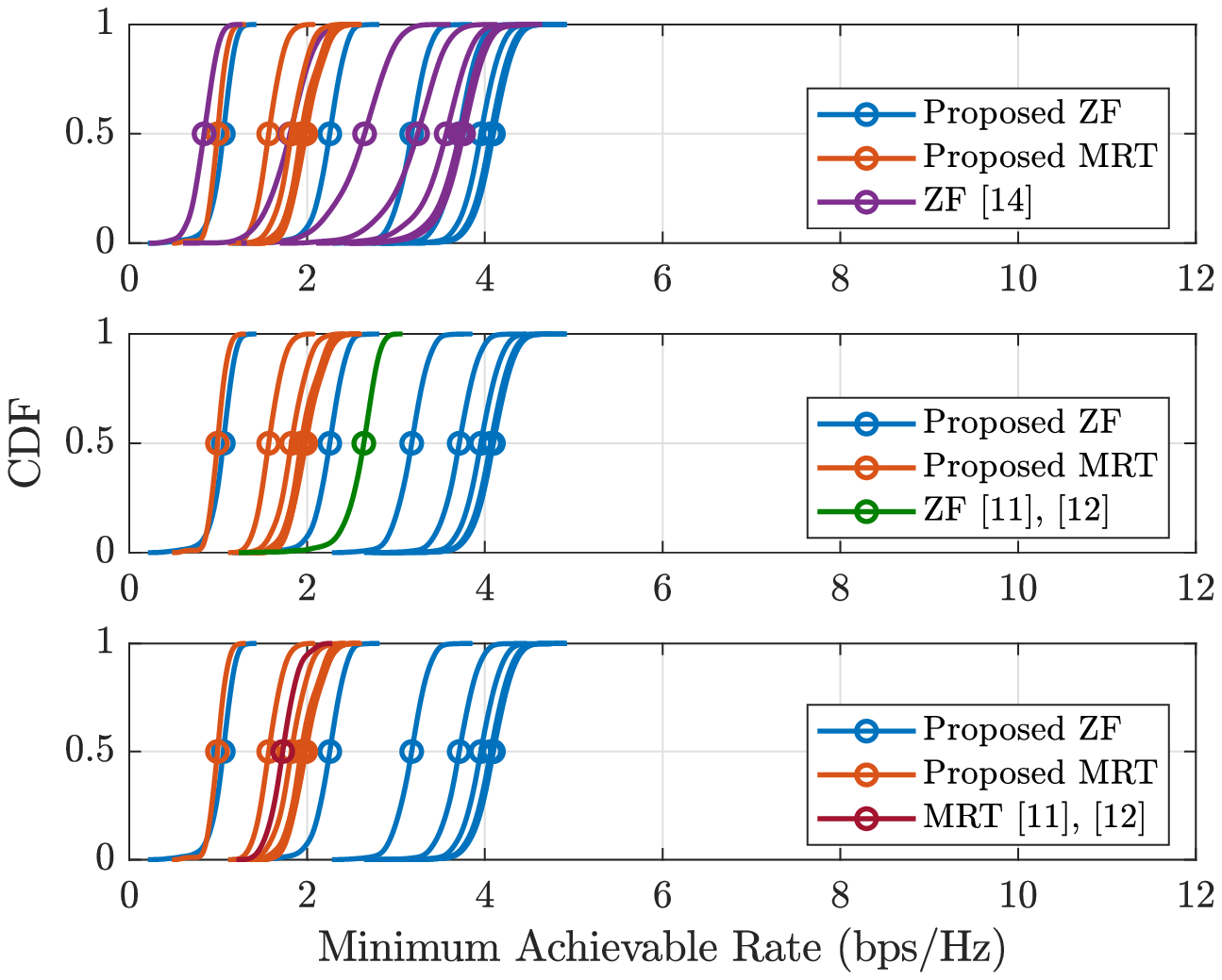}
\caption[caption]{The CDF of the minimum achievable rate lower bound of the users for $C=16$ bps/Hz. In each subplot, the lines with the same color correspond to $B=1, \dots, 8$ bits from left to right. The MRT and ZF precoders proposed in \cite{8756265, 9123382} are constrained to $C/(2R)=B=2$.}\label{figure_4}
\end{figure}

\begin{figure}[t]
\centering
\includegraphics[width=1\columnwidth]{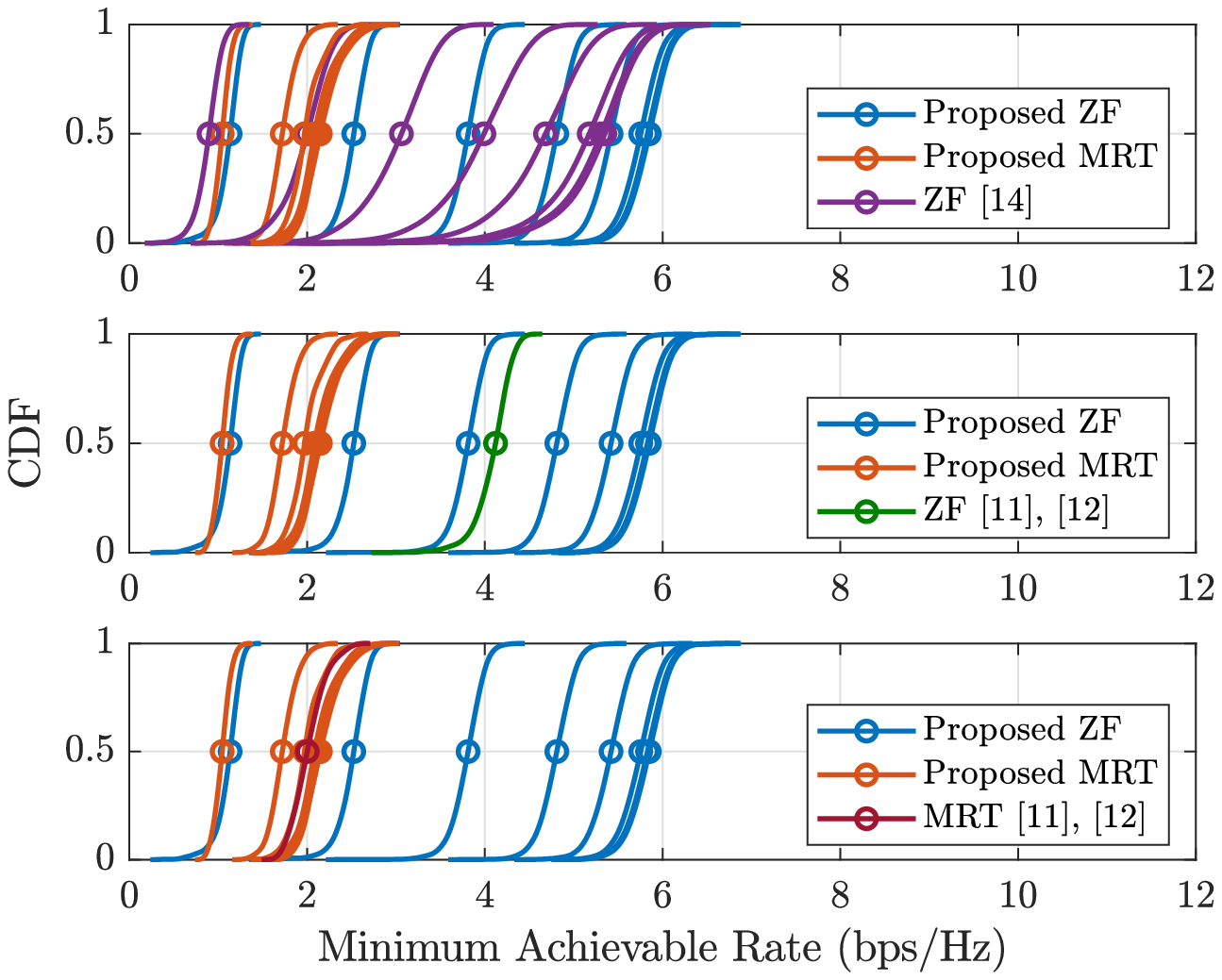}
\caption[caption]{The CDF of the minimum achievable rate lower bound of the users for $C=24$ bps/Hz. In each subplot, the lines with the same color correspond to $B=1, \dots, 8$ bits from left to right. The MRT and ZF precoders proposed in \cite{8756265, 9123382} are constrained to $C/(2R)=B=3$.}\label{figure_5}
\end{figure}

\begin{figure}[t]
\centering
\includegraphics[width=1\columnwidth]{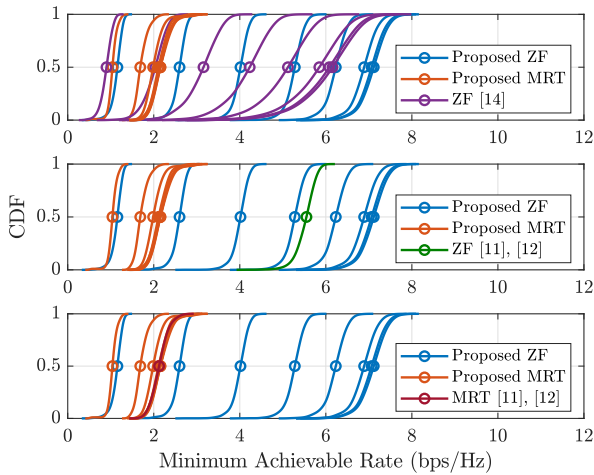}
\caption[caption]{The CDF of the minimum achievable rate lower bound of the users for $C=32$ bps/Hz. In each subplot, the lines with the same color correspond to $B=1, \dots, 8$ bits from left to right. The MRT and ZF precoders proposed in \cite{8756265, 9123382} are constrained to $C/(2R)=B=4$.}\label{figure_6}
\end{figure}

In the second simulation, we explore the cumulative distribution function (CDF) of the minimum achievable rate lower bound of the users for $C=8, 16, 24, 32$ bps/Hz and $B=1, \dots, 8$ bits where Figs. \ref{figure_3}-\ref{figure_6} correspond to $C=8, 16, 24, 32$ bps/Hz each containing $B=1, \dots, 8$ bits. According to Figs. \ref{figure_3}-\ref{figure_6}, Algorithm \ref{algorithm_2} outperforms Algorithm \ref{algorithm_1}, which is expected because the ZF precoder outperforms the MRT precoder in cell-free massive MIMO as observed in \cite{7827017, 7917284}. Also, Algorithm \ref{algorithm_2} yields higher rates than the heuristic approach of \cite{8678745} that aims to solve \eqref{p7}. The main shortcoming of the solution proposed in \cite{8678745} is that some of the constraints in \eqref{p7} are discarded. In contrast, Algorithm \ref{algorithm_2} accounts for all the constraints, thereby converging to the local optimum. From the first subplot in Fig. \ref{figure_6}, note that the 5\%-outage capacity of Algorithm \ref{algorithm_2} is higher than that of \cite{8678745} by 37\% for $B=8$ bits. Lastly, note that quantizer-based cell-free massive MIMO \cite{8756265, 9123382} is constrained to $C/(2R)=B$, which limits the system design flexibility. In contrast, our system model can attain higher rates by increasing the ADC/DAC resolution for fixed $C$ as illustrated in Figs. \ref{figure_3}-\ref{figure_6}. As a result, our system model with Algorithm \ref{algorithm_2} for $B=8$ bits yields a 5\%-outage capacity higher than that of \cite{8756265, 9123382} by 28\% as evident from the second subplot in Fig. \ref{figure_6}.

Before moving on, we illustrate the advantage of the MRT precoder over the ZF precoder in the implementation point of view. To compute the ZF precoder in cell-free massive MIMO, all the channel estimates must be available to the control unit as evident from \eqref{pinv}. In contrast, the MRT precoder can be computed in a decentralized fashion across the base stations as evident from the fact that the MRT precoder is given by $\{\mathbf{f}_{m, k}\}_{\forall m, k}=\{\hat{\mathbf{g}}_{m, k}\}_{\forall m, k}$. Therefore, the MRT precoder is inferior to the ZF precoder, but the implementation requirement is more flexible in the sense that the MRT precoder can be computed in a decentralized fashion unlike the ZF precoder. The interested reader is referred to \cite{8845768} for a more thorough discussion on how the implementation of cell-free massive MIMO can be classified to four different cooperation levels.

\begin{figure}[t]
\centering
\includegraphics[width=1\columnwidth]{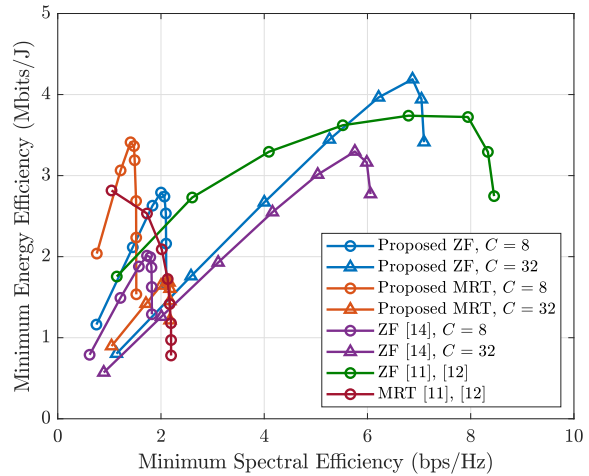}
\caption[caption]{The minimum energy-spectral efficiency lower bound as a function of $B$ for $C=8, 32$ bps/Hz. At each line, the markers correspond to $B=1, \dots, 8$ bits from left to right. The MRT and ZF precoders proposed in \cite{8756265, 9123382} are constrained to $C/(2R)=B=1, \dots, 8$.}\label{figure_7}
\end{figure}

\begin{figure}[t]
\centering
\includegraphics[width=1\columnwidth]{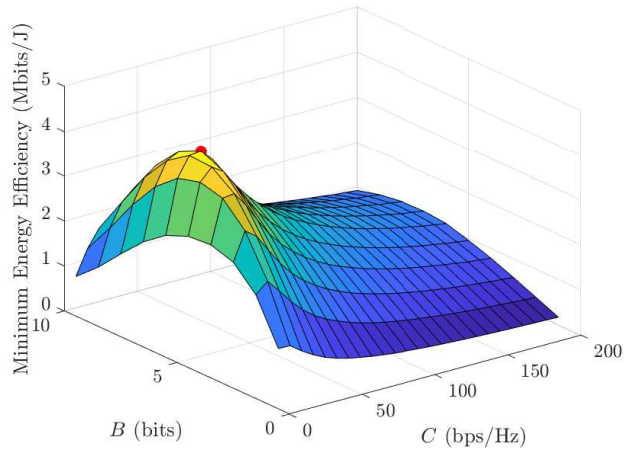}
\caption[caption]{The minimum energy efficiency lower bound of the ZF precoder for $C=8, 16, 24, \dots, 200$ bps/Hz and $B=1, \dots, 10$ bits. The red point corresponds to the most energy-efficient operating point that attains $\mathrm{EE}=4.19$ Mbits/J at $C=32$ bps/Hz and $B=6$ bits.}\label{figure_8}
\end{figure}

In the third simulation, we investigate the minimum energy efficiency lower bound. Fig. \ref{figure_7} depicts the minimum energy-spectral efficiency as a function of $B=1, \dots, 8$ bits for $C=8, 32$ bps/Hz. As shown in Fig. \ref{figure_7}, Algorithm \ref{algorithm_2} attains the highest energy efficiency among all the schemes, which verifies the superior performance of our scheme. In particular, the peak energy efficiency of Algorithm \ref{algorithm_2} is higher than that of the best baseline by approximately 15\%. There are some points outperformed by the ZF precoder of \cite{8756265, 9123382} in terms of spectral efficiency, but those are achieved by employing infinite-capacity fronthaul links in conjunction with Algorithm \ref{algorithm_2}. We also note that there is a trade-off between energy and spectral efficiency as a function of $B$ as energy efficiency hits the ceiling and eventually levels off. Since energy efficiency depends on $C$ and $B$ in an intertwined manner, the parameter that maximizes energy efficiency is searched exhaustively as illustrated in Fig. \ref{figure_8} for the ZF precoder. The most energy-efficient operating point is $C=32$ bps/Hz and $B=6$ bits that attains $\mathrm{EE}=4.19$ Mbits/J for the scenario of interest. These numerical studies provide guidelines onto how to optimize energy efficiency for future wireless communications.

\section{Conclusion}\label{section_5}
We considered the uplink channel estimation and downlink precoding in cell-free mmWave massive MIMO systems with low-capacity fronthaul links and low-resolution ADC/DACs. To model the nonlinear distortion from the fronthaul compression and ADC/DAC quantization, the information theoretic argument and AQNM were adopted. To minimize the channel estimation error, the codebook associated with the fronthaul compression was optimized. In the data transmission phase, the max-min fairness problem was addressed for the MRT and ZF precoders. The optimization was performed over the power coefficient and codebook parameter based on the proposed schemes in an alternating fashion. The simulation results showed that the proposed schemes outperform several state-of-the-art baselines in terms of energy and spectral efficiency.

\bibliographystyle{IEEEtran}
\bibliography{refs_all}

\end{document}